\documentclass[%
   reprint,
  showpacs,preprintnumbers,
  amsmath,amssymb,
]{revtex4-1}

  \usepackage{graphicx}
  \usepackage{dcolumn}
  \usepackage{bm}
  \usepackage{color}
  \usepackage{multirow}
  \usepackage{esint}
  \usepackage{acro}
  
  \makeatletter
  \newcommand*{\addFileDependency}[1]{
    \typeout{(#1)}
    \@addtofilelist{#1}
    \IfFileExists{#1}{}{\typeout{No file #1.}}
  }
  \makeatother

  \newcommand*{\myexternaldocument}[1]{%
    \externaldocument{#1}%
    \addFileDependency{#1.tex}%
    \addFileDependency{#1.aux}%
  }
  \usepackage{xr}
  \myexternaldocument{suppl_mat}
  \DeclareAcronym{WP}{
    short=WP,
    long=Weyl point,
    }
  \DeclareAcronym{DOS}{
    short=DOS,
    long=density of states,
    }
  \DeclareAcronym{bct}{
    short=bct,
    long=body-centered tetragonal,
    }
  \DeclareAcronym{ARPES}{
    short=ARPES,
    long=angle-resolved photoemission spectroscopy,
    }
  \DeclareAcronym{SOC}{
    short=SOC,
    long=spin-orbit coupling,
    }
  \DeclareAcronym{3D}{
    short=3D,
    long=three-dimensional,
    }
  \DeclareAcronym{2D}{
    short=2D,
    long=two-dimensional,
    }
  \DeclareAcronym{DFT}{
    short=DFT,
    long=density functional theory,
    }
  \DeclareAcronym{XC}{
    short=XC,
    long=exchange and correlation,
    }
  \DeclareAcronym{BZ}{
    short=BZ,
    long=Brillouin zone,
    }
  \DeclareAcronym{TWS}{
    short=TWS,
    long=topological Weyl semimetal
    }
  \DeclareAcronym{TDS}{
    short=TDS,
    long=topological Dirac semimetal
    }
  \DeclareAcronym{PBE}{
    short=PBE,
    long={Perdew, Burke and Ernzerhof},
    }
  \DeclareAcronym{GGA}{
    short=GGA,
    long=generalized gradient approximation,
    }
  \DeclareAcronym{PW}{
    short=PW,
    long=plane wave,
    }
  \DeclareAcronym{EELS}{
    short=EELS,
    long=electron energy loss spectra,
    }
  \DeclareAcronym{QE}{
    short=QE,
    long=QUANTUM ESPRESSO,
    }
  \DeclareAcronym{IR}{
    short=IR,
    long=infrared,
    }

\begin{document}
  \title{Influence of anisotropy, tilt and pairing of Weyl nodes: \\The Weyl semimetals TaAs, TaP, NbAs and NbP}

  \author{Davide Grassano, Olivia Pulci, Elena Cannuccia}
  \affiliation{%
    Dipartimento di Fisica, Universit\`a "Tor Vergata" Roma and INFN, via della Ricerca Scientifica, I-00133 Rome, Italy
    }%

  \author{Friedhelm Bechstedt}
  \affiliation{%
    Institut f\"ur Festk\"orpertheorie und -optik, Friedrich-Schiller-Universit\"at Jena,\\ Max-Wien-Platz 1, 07743 Jena, Germany
    }%

  \date{\today}

  \begin{abstract}
    By means of \emph{ab initio} band structure methods and model Hamiltonians we investigate the electronic, spin and topological properties of four monopnictides crystallizing in \acl{bct} structure.
    We show that the Weyl bands around a \acl{WP} W1 or W2 possess a strong anisotropy and tilt of the accompanying Dirac cones.
    These effects are larger for W2 nodes than for W1 ones.
    The node tilts and positions in energy space significantly influence the \acl{DOS} of single-particle Weyl excitations.
    The node anisotropies destroy the conventional picture of (anti)parallel spin and wave vector of a Weyl fermion.
    This also holds for the Berry curvature around a node, while the monopole charges are independent as integrated quantities.
    The pairing of the nodes strongly modify the spin texture and the Berry curvature for wave vectors in between the two nodes.
    Spin components may change their orientation.
    Integrals over planes perpendicular to the connection line yield finite Zak phases and winding numbers for planes between the two nodes, thereby indicating the topological character.
  \end{abstract}
  \acresetall

  \maketitle

  \section{\label{sec1}Introduction}
    All fermions discovered in the menagerie of conventional particles have a finite mass.
    However, already in 1929 Weyl \cite{Weyl:1929:ZfP} conjectured that massless relativistic fermions may exist and can be described by a two-component spinor in contrast to the four-component Dirac spinor \cite{Dirac:1928:PRSLAMa}.
    In a many-electron system, electronic excitations can be described by quasiparticles \cite{Bechstedt:2015:Book}, where electrons collectively behave as if they are one elementary particle, for instance a Weyl fermion.
    The existence of Weyl fermions as low-energy electronic excitations was predicted by Herring \cite{Herring:1937:PR} in the context of electronic band structures of crystals.
    Indeed, such massless quasiparticles have been recently discovered in the so-called Weyl semimetals \cite{Burkov:2016:NM} of the TaAs compound family \cite{Weng.Fang.ea:2015:PRX,Huang.Xu.ea:2015:NC,Xu.Belopolski.ea:2015:S,Lv.Weng.ea:2015:PRX,Yang.Liu.ea:2015:NP}.
    They have been identified by topological Fermi arcs on the surface, Dirac cones (also called Weyl cones) formed by bulk linear Bloch bands, and chiral magnetic effects in the bulk \cite{Jia.Xu.ea:2016:NM}.
    The fine tuning of their topological properties have been demonstrated, in particular by \ac{ARPES} \cite{Liu.Yang.ea:2015:NM,Xu.Belopolski.ea:2016:PRL}.

    In such crystals Weyl fermions exist as low-energy excitations in their fully relativistic band structure.
    Scalar relativistic mass-Darwin effects in the electron shells of heavy elements lead to an energetic overlap and an inverted ordering of $s$, $p$, or $d$ orbitals.
    Strong \ac{SOC}, however, makes the bulk bands to be gapped in the momentum space except at some isolated linearly crossing points ${\bf k}_W$, named Weyl points.
    In contrast to a Weyl (or Dirac for higher symmetries) semimetal, in topological insulators \ac{SOC} opens a finite gap but keeps the band inversion \cite{Yan.Felser:2017:ARCMP}.
    The zero-gap semiconductors HgTe and $\alpha$-Sn are striking examples for band inversion due to the scalar-relativistic and spin-orbit effects \cite{Kuefner.Bechstedt:2015:PRB}.
    Although such band touching has long been known\cite{Herring:1937:PR}, its topological nature has been discovered only recently \cite{Wan.Turner.ea:2011:PRB}.

    The massless fermions discovered as quasiparticles in crystals, which are characterized by conduction and valence bands locally forming Dirac cones, are similar to those found for \ac{2D} graphene \cite{Burkov:2016:NM} and proximal to the Fermi energy.
    In \ac{3D} Dirac semimetals such as Cd$_3$As$_2$ with both inversion and time-reversal symmetry, the bands around the Dirac touching points form \ac{3D} Dirac cones \cite{MoscaConte.Pulci.ea:2017:SR}.
    The linear bands are due to As $p$ orbitals, oriented toward the \ac{3D} screw arrangement of the Cd vacancies.
    In \ac{3D} Weyl semimetals, the lower symmetry leads to a lifting of the fourfold degeneracy at the Dirac points, while the band structure remains gapless.
    However, in contrast to the \ac{3D} Dirac and Weyl semimetals, the \ac{2D} graphene-like honeycomb crystals show a \ac{SOC}-induced gap opening between the upper and lower Dirac cones \cite{Matthes.Pulci.ea:2013:JoPCM,grassano2018detection}.

    There is a third distinction between \ac{3D} graphene analogues and \ac{2D} graphene.
    In general, the upper and lower Dirac cones in \ac{3D} Dirac and Weyl semimetals are tilted and anisotropic \cite{Burkov:2016:NM,Xu.Belopolski.ea:2015:S,Lv.Weng.ea:2015:PRX,MoscaConte.Pulci.ea:2017:SR,Lee.Xu.ea:2015:PRB,Souma.Wang.ea:2016:PRB,Xu.Autes.ea:2017:PRL,Grassano.Bechstedt.ea:accepted}.
    Both deviations from the ideal Dirac cone picture have consequences, especially for the physical properties of Weyl semimetals.
    Crystals hosting Weyl fermions decay in two classes, type-I and type-II ones \cite{Soluyanov.Gresch.ea:2015:N}, in dependence on the tilt strength.
    Such a strong tilt also breaks the Lorentz invariance of Weyl fermions in quantum field theory.
    It may be important also for the strength of photocurrents for optical excitations between the lower and upper Dirac cones \cite{Chan.Lindner.ea:2017:PRB}.
    The anisotropy of the Dirac cones touching at a Weyl point influences several properties like optical \cite{Grassano.Bechstedt.ea:accepted}, spin texture \cite{Xu.Belopolski.ea:2016:PRL}, and transport \cite{Shekhar.Nayak.ea:2015:NP,Zhang.Xu.ea:2016:NC} ones.
    In a non-centrosymmetric Weyl semimetal the Weyl nodes appear in pairs of opposite tilting, rotated anisotropy, and of different chirality, either left-handed or right-handed.
    In the symmetric limit they act as sources and sinks of Berry curvature \cite{Wan.Turner.ea:2011:PRB}.
    Therefore, such Weyl semimetals give rise to a chiral anomaly in quantum transport \cite{Zhang.Xu.ea:2016:NC}, which also leads to fingerprints in surface spectroscopies \cite{Mitchell.Fritz:2016:PRB,Behrends.Grushin.ea:2016:PRB}.
    The chiral anomaly, the first known example for a quantum anomaly, refers to the nonconservation of a chiral current, i.e., a current imbalance between two distinct species of chiral fermions and, hence, Weyl nodes in a pair.
    In the case of real Weyl semimetals with tilted, anisotropic and paired Dirac cones, the energy restriction of Weyl fermion excitations has to be investigated \cite{Grassano.Bechstedt.ea:accepted}.

    In this paper we theoretically investigate the anisotropy and tilt of Dirac cones at Weyl points in non-centrosymmetric Weyl semimetals by a combination of \emph{ab initio} band structure calculations and model Hamiltonians for each Weyl point, whose parameters are fitted to the band-structure results.
    The transition metal monopnictides TaAs, TaP, NbAs and NbP crystallizing in a \ac{bct} structure are considered as model Weyl systems.
    The methodology is described in Sec.~\ref{sec2}.
    In Sec.~\ref{sec3} the anisotropy and the tilt of the Dirac cones in pairs of Weyl points of two different types are characterized by tilt vectors and tensors of Fermi velocities.
    In the following sections~\ref{sec4} and \ref{sec5} the consequences for spin textures and topological properties, respectively, are discussed.
    Finally, a summary and conclusions are given in Sec.~\ref{sec6}.

  \section{\label{sec2}Methodology}
    \subsection{\label{sec2a}\emph{Ab initio} calculations}
      The structural, electronic and topological properties of the model monopnictides are investigated in the framework of the \ac{DFT} as implemented in the Quantum Espresso \cite{Giannozzi.Baroni.ea:2009:JoPCM,espresso2017} package.
      \Ac{XC} are treated within the generalized gradient approximation of Perdew, Burke, and Ernzerhof \cite{Perdew.Burke.ea:1996}.
      The electron-ion interaction is described by norm-conserving, fully relativistic pseudopotentials.
      The $s$ and $p$ valence electrons are generally taken into account.
      For the Ta and Nb atoms also the partially filled semicore $d$ shells are considered, for Ta even the filled Ta4$f$ one.
      More in detail, we use pseudoatoms with the electronic configurations 4$s^24p^64d^35s^2$ (Nb), $4f^{14}5s^25p^65d^36s^2$ (Ta), $4s^24p^3$ (As), and $3s^23p^3$ (P) down to binding energies of the order of 40~eV.
      Relativistic effects are fully taken into account.
      Besides the scalar-relativistic mass correction and Darwin term, the \ac{SOC} is also included in the pseudopotentials.
      The details are described in Ref. \cite{Grassano.Bechstedt.ea:accepted}.
      The correspondingly optimized atomic geometries of the four \ac{bct} crystals with the non-symmorphic space group symmetry I$4_1md$ ($C^1_{4v}$) are used.

      Also the same electronic structure calculations are applied to the electronic bands and wave functions.
      Since we focus on the low-energy Weyl fermion excitations around the Fermi level, we neglect quasiparticle effects \cite{Bechstedt:2015:Book}, which, in \ac{2D} graphene-like crystals, only increase the Fermi velocities by about $10-20$~\% \cite{Matthes.Pulci.ea:2013:JoPCM,Yang.Deslippe.ea:2009:PRL}.
      This tendency should be weakened in \ac{3D} semimetals because of the reduced screening due to higher dimensions and the presence of free carriers.
      Consequently, we use the Kohn-Sham eigenvalues $\varepsilon_\nu({\bf k})$ and eigenstates $|\nu{\bf k}\rangle$ of the \ac{DFT} for each Bloch state with band index $\nu$ and wave vector ${\bf k}$ to investigate the electronic structure and related properties.

    \subsection{\label{sec2b}Model Hamiltonian: One tilted anisotropic node}
      We linearize the ${\bf k}$-dispersion $\varepsilon_\nu({\bf k})$ near a single \ac{WP} ${\bf k}_W$ in the \ac{bct} \ac{BZ}, i.e., in ${\bf q}={\bf k}-{\bf k}_W$, where ${\bf q}=0$ defines the valence and conduction band touching point.
      Following the ideas of the ${\bf k}\cdot{\bf p}$ theory, this allows us to formulate a stationary Weyl equation for the corresponding single Weyl fermion excitations in real space by
      \begin{align}\label{eq1}
        \hat{H}_W \Psi_{\nu{\bf q}}({ \bf x}) =
        \varepsilon_\nu({\bf q})
        \Psi_{\nu{\bf q}}({\bf x})
      \end{align}
      with $(\nu=\pm)$ for the upper(lower) Weyl bands and the generalized Weyl Hamiltonian \cite{Burkov:2016:NM,Weng.Fang.ea:2015:PRX,Wan.Turner.ea:2011:PRB,Soluyanov.Gresch.ea:2015:N,Hosur.Qi:2013:CRP}
      \begin{align}\label{eq2}
        \hat{H}_W =
        ({\bf v}_0{\bf p})\hat
        {\sigma}_0 +
        {\bf p}
        \left(
          \hat{v}_F \hat{\bm\sigma}
        \right)
        ,
      \end{align}
      where $\hat{\sigma}_0$ is the 2$\times$2 unit matrix, $\hat{\bm\sigma}=(\hat{\sigma}_x,\hat{\sigma}_y,\hat{\sigma}_z)$ is the vector of the three 2$\times$2 Pauli matrices, ${\bf p}$ is the quantum-mechanical momentum operator, ${\bf v}_0$ denotes the tilt velocity vector, and
      \begin{align}\label{eq3}
        \hat{v}_F=\left(
        \begin{array}{ccc}
        v_{xx} & v_{xy} & v_{xz} \\
        v_{xy} & v_{yy} & v_{yz} \\
        v_{xz} & v_{yz} & v_{zz}
        \end{array}
        \right)
      \end{align}
      represents the symmetric matrix of the Fermi velocities, which characterize the anisotropy of the upper and lower Dirac cones.

      The eigenvalues of \eqref{eq1} are 
      \begin{align}\label{eq4}
        \begin{split}
          \varepsilon_\nu({\bf q})= & T({\bf q})\pm U({\bf q}), \\
          T({\bf q})= & \hbar {\bf v}_0{\bf q},
          \\
          U({\bf q})= & \hbar
          \left[
            \sum_{j=x,y,z}
              \left(
                {\bf v}_j{\bf q}
              \right)^2
          \right]^{\frac{1}{2}} = 
          \hbar
          |
            \hat{v}_{F}{\bf q}
          |
        \end{split}
      \end{align}
      with ${\bf v}_j$ as a vector representing a column or a row of the matrix $\hat{v}_F$ of Fermi velocities \eqref{eq3}.
      One can assign a chirality (or chiral charge) $c=\pm1$ to the fermions defined as $c={\rm sgn}({\bf v}_x \cdot ({\bf v}_y \wedge {\bf v}_z))={\rm sgn}(det(\hat{v}_F))$ \cite{Wan.Turner.ea:2011:PRB}.
      Taking into account the chirality one may generalize the result \eqref{eq4} by
      $\varepsilon_\pm({\bf k} - {\bf k}_W)
        \Rightarrow
      c \varepsilon_\pm({\bf k} - c {\bf k}_W)$, thereby relating the Weyl bands at the paired Weyl points to each other.

      The orthonormalized eigenvectors are Weyl-Pauli spinors of the type
      \begin{align}\label{eq5}
        \begin{split}
          \Psi_{\nu{\bf q}}( {\bf x}) = &
          U_{\nu{\bf q}}
          \frac
            {1}
            {\sqrt{V}}
          e^{ i{\bf qx}} ,
          \\
          U_{\nu{\bf q}} = &
          \left(
            \begin{array}{c}
              u^u_{\nu{\bf q}} \\
              u^l_{\nu{\bf q}}
            \end{array}
          \right),
        \end{split}
      \end{align}
      where the upper (lower) part $u^u_{\nu{\bf q}}$ ($u^l_{\nu{\bf q}}$) of the spinor represents some Bloch factor near ${\bf q}=0$.
      The plane-wave part is normalized to the crystal volume $V$.
      The Bloch factors are
      \begin{align}\label{eq6}
        \begin{split}
          u^u_{\pm {\bf q}} = &
          \frac
            {1}
            {\sqrt{2}}
          \sqrt{
            1 \pm
            f({\bf q})
          }
          e^{ -i \theta({\bf q}) / 2}
          ,
          \\
          u^l_{\pm {\bf q}} = & 
          \pm \frac
            {1}
            {\sqrt{2}}
          \sqrt{
          1 \mp 
          f({\bf q})
          }
          e^{ i \theta({\bf q}) / 2}
        \end{split}
      \end{align}
      with
      \begin{align}\label{eq7}
        \begin{split}
          f({\bf q}) = &
          \frac
          {({\bf v}_z {\bf q})}
          {U({\bf q}) / \hbar}
          ,
          \\
          \theta({\bf q}) = &
          \arctan
          \frac
            {({\bf v}_y {\bf q})}
            {({\bf v}_x {\bf q})}
          .
        \end{split}
      \end{align}
      Due to the degeneracy of the two bands $\nu=+,-$ at ${\bf q}=0$, the \ac{WP} ${\bf q}=0$ is a non-analytical point.
      The wave functions and all properties related to matrix elements depend for ${\bf q} \rightarrow 0$ on the direction ${\bf q}/q$ of the wave vector.
      Because of the two-component spinors \eqref{eq5} the Weyl particles are two-component fermions.
      The only way for them to disappear is if they meet with another two-component Weyl fermion in the \ac{BZ}, with opposite chiral charge.
      Thus, they are topological objects.
      This may happen for the second Weyl node at ${\bf k^{\prime}}_W$ in the pair, which has the opposite chirality, since the Fermi velocity vectors are reversed \cite{Wan.Turner.ea:2011:PRB}.

    \subsection{\label{sec2c}Model Hamiltonian:  pairing of nodes}
      The electronic properties around a pair of Weyl nodes, separated along the $k_x$-axis at $-k_{Wx}$ and $k_{Wx}$, may be described within a minimal (without tilting and off-diagonal elements) two-node model \cite{Lu.Zhang.ea:2015:PRB,Zhang.Lu.ea:2016:NJoP,Lu.Shen:2017:FoP}
      \begin{align}\label{eq8}
        \hat{H}_{2W}=v_{yy}(p_y\hat{\sigma}_x+p_z\hat{\sigma}_y)-\frac{1}{2}\frac{v_{xx}}{k_{Wx}}(k^2_{Wx}-{\bf p}^2)\hat{\sigma}_z.
      \end{align}
      The sign of the second term has been chosen to approach to \eqref{eq2} for small wave vector deviations from the right \ac{WP} in the pair.
      The eigenvalues and eigenfunctions of the corresponding Weyl equation \eqref{eq1} are classified by a wave vector ${\bf Q}=(k_x,q_y,q_z)$, where $k_x$ runs in the surroundings of both Weyl points $(\pm k_{Wx},k_{Wy},k_{Wz})$ whereas $q_y=k_y-k_{Wy}$ and $q_z=k_z-k_{Wz}$ represent reduced wave vectors.
      Again, we find two Weyl bands
      \begin{align}\label{eq9}
        \varepsilon_\pm( {\bf Q}) = 
        \pm \hbar
        \left[
          v^2_{yy}
          \left(
            q^2_y +
            q^2_z
          \right) +
          \left(
            \frac
              {v_{xx}}
              {2k_{Wx}}
          \right)^2
          ( k^2_{Wx} - Q^2)^2
        \right]^{\frac{1}{2}}
      \end{align}
      with the same structure of the Weyl-Pauli spinors $\Psi_{\pm {\bf Q}}({\bf x})$ as described in \eqref{eq5} and \eqref{eq6} but with a modification of \eqref{eq7},
      \begin{align}\label{eq10}
        \begin{split}
          \theta({\bf Q}) = &
          \arctan( q_z / q_y),
          \\
          f({\bf Q}) = &
          \frac
            {
              -
              \frac
                  {v_{xx}}
                  {2 k_{Wx}}
              ( k^2_{Wx} - Q^2)
            }
            {
              \left[
                \left(
                  \frac
                    {v_{xx}}
                    {2k_{Wx}}
                \right)^2
                \left(
                  k^2_{Wx} -
                  Q^2
                \right)^2 +
                v^2_{yy}
                ( q^2_y + q^2_z)
              \right]^{\frac{1}{2}}}
          .
        \end{split}
      \end{align}
      Close to the two Weyl points, with $q_x=k_x-k_{Wx}$ and $q_x=k_x+k_{Wx}$, one has
      \begin{align}\label{eq11}
        \varepsilon_\pm( {\bf Q}) = 
        \pm \hbar 
        \left[
          v^2_{yy}
          (q^2_y + q^2_z) +
          v^2_{xx} q^2_x
        \right]^{\frac{1}{2}}
      \end{align}
      and
      \begin{align}\label{eq12}
        f( {\bf Q}) = 
        \pm
        \frac
          {v_{xx}q_x}
          {
            \left[
              v^2_{yy}
              \left(
                q^2_y +
                q^2_z
              \right) + 
              v^2_{xx} q^2_x
            \right]^{\frac{1}{2}}
          }
      \end{align}
      for the WP at $+k_{Wx}$, while the signs have to be reversed for the \ac{WP} at $-k_{Wx}$.
      The results for the right isolated Weyl point agree with those in \eqref{eq4}--\eqref{eq7} for vanishing tilt, i.e., ${\bf v}_0=0$, and a diagonal matrix of the Fermi velocities \eqref{eq3}
      \begin{align}\label{eq13}
        \hat{v}_F=
        \left(
        \begin{array}{ccc}
          v_{xx} & 0 & 0 \\
          0 & v_{yy} & 0 \\
          0 & 0 & v_{yy}
        \end{array}
        \right),
      \end{align}
      characterizing a reduced anisotropy due to the pairing of WPs.
      For the left Weyl point the sign of $v_{xx}$ has to be changed in \eqref{eq13}.
      We have to point out, that compared to Sec.~\ref{sec2b}, the local coordinate systems has been changed from ${\bf Q} = (k_x,q_y,q_z)$ to ${\bf q} = ( q_x, q_y, q_z)$.

  \section{\label{sec3}Tilt and Anisotropy of Weyl Cones}
    \subsection{\label{sec3a}Weyl nodes}
      The band structures of the four \ac{bct} transition metal monopnictides TaAs, TaP, NbAs, and NbP in Fig.~SM6 are very similar.
      The main differences are due to the reduction of \ac{SOC} going from TaX to NbX (X=As, P).
      Four pairs of W1 Weyl points and eight pairs of W2 ones appear in the  \ac{BZ} \cite{Weng.Fang.ea:2015:PRX}.
      Their positions can be hardly identified in the band structures along high-symmetry lines.
      Some indications seem to be visible near a $\Sigma$ point and the $\Sigma'N$ high-symmetry line.
      However, the band crossings near the $Z\Sigma'$ line (near $Z$) are gapped and, therefore, indicate a trivial behavior.
      The corresponding hole pockets do not give rise to Weyl fermions but support the semimetallic character.

      The ${\bf k}$-space location of the 24 Weyl points are determined by means of extremely dense ${\bf k}$-point grids \cite{Grassano.Bechstedt.ea:accepted}.
      An illustration of the distribution of the 12 pairs of WPs is given in Fig.~SM1 for TaAs while an indication of the ${\bf k}$-space positions of trivial hole pockets can be found in Fig.~SM2.
      The resulting 24 positions ${\bf k}_W$ are listed in Table~SM1 for the four Weyl semimetals.
      
      Despite their influence on transport and optical measurements, the trivial points, i.e., the positions of hole (or electron) pockets without band crossing, will be discussed more indirectly here, because they do not give rise to Weyl excitations, but influence the Fermi level position.
      The high-density {\bf k}-point sampling of the  \ac{BZ}, used to find the Weyl points, allow us also to identify many trivial points at non-high-symmetry positions with an increasing number from TaAs to NbP.
      In particular, the phosphides possess more of such points.
      We found, at least, 48 (108, 100, 175) trivial points in the \ac{BZ} for TaAs (TaP, NbAs, NbP).
      They are located in the $k_x = 0$ and $k_y =0$ planes, especially near the zone boundaries (for TaAs see Fig.~SM1).
      Thereby, their $k_z$ values vary in the range of intermediate values measured in terms of the \ac{BZ} extent.
      They form clusters near the four edges of the \ac{BZ}, where $k_x$ and $k_y$ penetrate the \ac{BZ} surface. 

    \subsection{\label{sec3b}Weyl excitations}
      In order to discuss the low-energy Weyl fermions, the Bloch bands are plotted (see  Fig.~SM7) around one pair of Weyl points in an energy range of the order of 100~meV around the Fermi level.
      We choose a pair of \ac{WP} such that the connection line lies along the $k_x$-direction.
      Therefore, the band crossings are visible at $k_{Wx}$ and $-k_{Wx}$.
      However, not only crossing bands, which can be linearized, but also trivial parabolic bands nearby (mainly conduction bands for W1, valence bands for W2) are displayed.

      The calculated bands  clearly indicate that a linear behavior is only valid for small energy and wave-vector intervals around a Weyl point.
      This holds especially for the band dispersion between WPs of one pair along the connection line, which characterizes one source of anisotropy.
      The small extents of the linear branches are related to the small ${\bf k}$-space splitting $2k_{Wx}$ of the two nodes.
      However, the anisotropy of the bands perpendicular to the connection line, in particular for W$_1$ lines, is clearly visible.
      Qualitatively the same band dispersion  have been observed in several \ac{ARPES} experiments for TaAs and NbP \cite{Xu.Belopolski.ea:2015:S,Lv.Weng.ea:2015:PRX,Xu.Autes.ea:2017:PRL,Belopolski.Xu.ea:2016:PRL} but also TaP \cite{Xu.Belopolski.ea:2015:SA}.

      The positions of the band touching points with respect to the Fermi level $\varepsilon_F$, $\Delta\varepsilon_{F_1}=-26$ (TaAs), $-55$ (TaP), $-33$ (NbAs), and $-56$~meV (NbP) for W1 as well as $\Delta\varepsilon_{F_2}=-13$ (TaAs), 21 (TaP), 4 (NbAs), and 26~meV (NbP) for W2 agree well with similarly calculated results \cite{Lee.Xu.ea:2015:PRB} but also experimental data.
      In general, the $\Delta\varepsilon_{F_j}$ ($j=1,2$) data indicate electron pockets for W1 and hole pockets for W2 \acp{WP} in TaP, NbAs, and NbP, while in TaAs only electron pockets occur. 
      Free carriers can also appear in bands with quadratic wave-vector dispersion e.g. in  the second conduction band of NbAs and NbP near W1 points  (Fig.~SM7).
      The presence of hole pockets near trivial points within the \ac{BZ} (see Fig.~SM1) induces  difficulties to extract free carrier densities in Weyl bands near \acp{WP} by means of transport measurements.

    \subsection{\label{sec3c}Fermi velocities in tilted cones and semimetal type}
      The linearities of the two Weyl bands near the touching points of W1 and W2 type  suggest a fit to the linear dispersion relations \eqref{eq4}.
      Thereby, the  kinetic component $T({\bf q})$ \cite{Soluyanov.Gresch.ea:2015:N} is dominated by the tilt velocity vector ${\bf v}_0$.
      The so-called potential component of the energy spectrum, $U({\bf q})$, characterizes the shape of the upper and lower cones by the matrix $\hat{v}_F$ \eqref{eq3}.
      To determine the nine parameters in the vector ${\bf v}_0$ and the matrix $\hat{v}_F$ we made a tensor fit to the \emph{ab initio} calculated Weyl bands in Fig.~SM7.
      The velocities resulting for an extremely narrow region around the Weyl nodes of one pair separated in $k_x$-direction are listed in Table~\ref{tab2}.

\begin{figure*}[h]
    \includegraphics[width=0.95\textwidth,height=0.95\textheight,keepaspectratio]{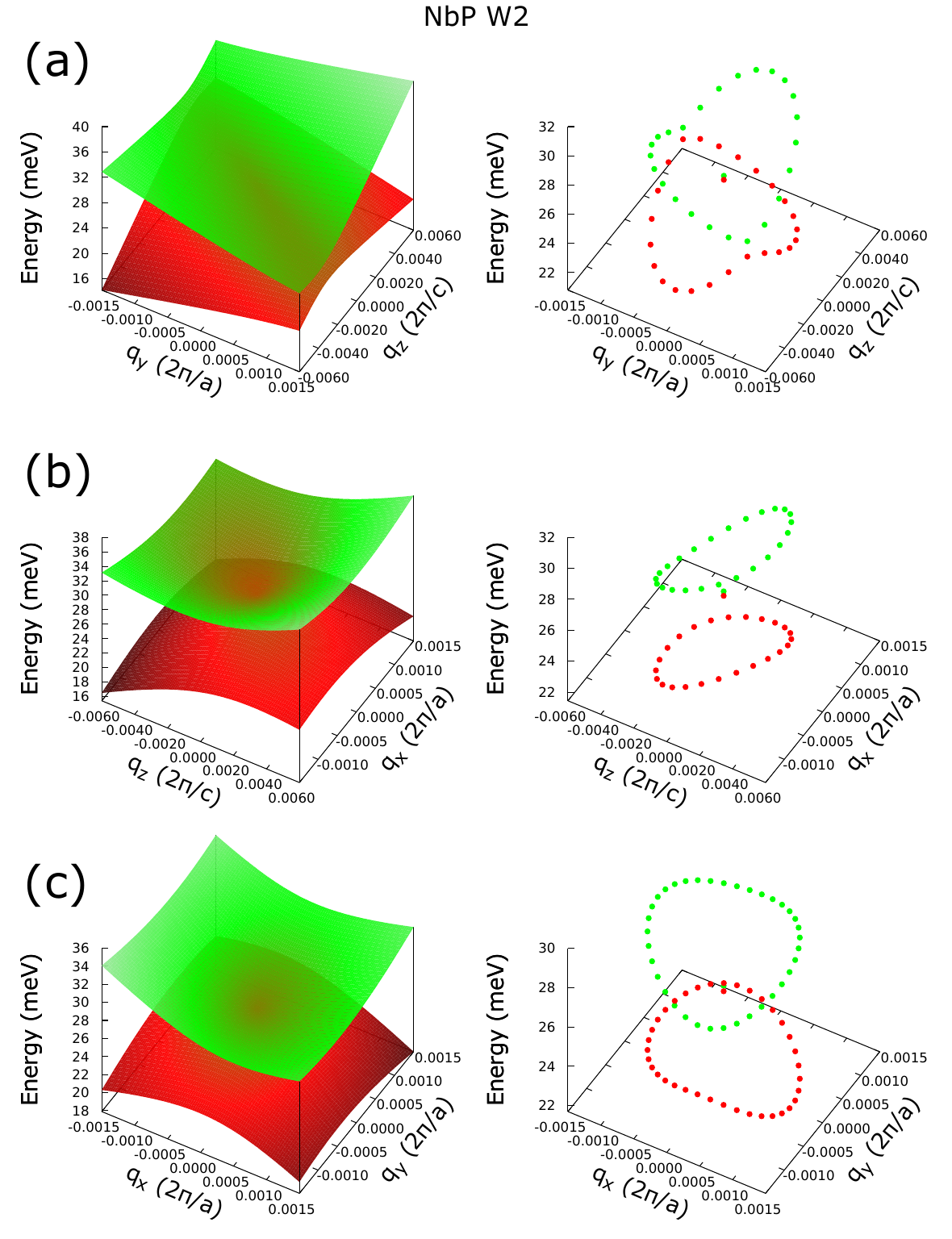}
    \caption{\label{fig:3D_fit}
      Two calculated Weyl bands near a W2 Weyl point for NbP.
      Three cuts, (a) $q_x=0 \ (k_x = k_{Wx})$, (b) $q_y=0$, and (c) $q_z=0$ plane, are displayed.
      Left panels: Plot of $\varepsilon_\nu ({\bf q})$ \eqref{eq4} using the fit parameter from Table~\ref{tab2}.
      Right panels: Plot of $\varepsilon_\nu ({\bf q})$ computed \emph{ab initio}.
      }
  \end{figure*}

      The velocities in Table~\ref{tab2} confirm the picture  of almost \ac{2D} W1 and \ac{3D} W2 Weyl bands (see Fig.~SM7).The tilt is illustrated in Fig.~\ref{fig:3D_fit} for W$_2$ in NbP.
      The resulting Fermi and tilt velocities support the picture highly anisotropic and strongly tilted Weyl cones.
      The order of magnitude of a few 10$^5$~m/s for the diagonal elements of the tensor $\hat{v}_F$ agrees with Fermi velocities in \ac{3D} but also \ac{2D} materials \cite{MoscaConte.Pulci.ea:2017:SR,Matthes.Pulci.ea:2013:JoPCM}, in general.
      However, they also suggest that our picture of Weyl fermions has to be generalized, because of the strong anisotropy and tilt. Whereas the off-diagonal elements are vanishing small for W$_1$, they are large, even of the order of magnitude of the diagonal velocities, in the W$_2$ case.
      The values in Table~\ref{tab2} indicate a complex Fermi surface, far away from a spherical shape. It decays into 24 contributions with the highestcomplexity around the W$_2$ Weyl nodes.

      The explicit values of the Fermi velocities $v_{ij}$ $(i,j=x,y,z)$ in Table~\ref{tab2} show clear trends and fulfill symmetry considerations: (i) They exhibit clear chemical trends with variations smaller than one order of magnitude.
      Usually the $v_{xx}$ value is the largest one, except for W2 in TaP.
      (ii) For W1 the off-diagonal elements are rather small, in particular in the case of the Nb compounds.
      The W2 \acp{WP} show a stronger velocity anisotropy because of the less symmetric positions ${\bf k}_{W_2}$ (see Table~SM1).
      (iii) Since the $k_x$-axis is chosen parallel to the connection line of two \acp{WP} in a pair and because of the $M_x$ symmetry, the matrix elements of the neighboring \acp{WP} change the sign if they contain at least one $x$ index.
      Consequently, the value $det(\hat{v}_F)$ changes also the sign indicating the opposite chiralities of the \acp{WP} in the pair, which survive the anisotropy. 
      (iv) The values of $det(\hat{v}_F)$ are negative for W2 and the Ta compounds.
      This is a consequence of the large off-diagonal elements of TaP.
      For instance, $v_{xz}$ is the largest matrix element.
      All tilt vectors have a large negative $x$-component.
      However,  they point in arbitrary directions, not only along the connection line of the two WPs in the chosen pair.

      The tilt ${\bf v}_0$ of the anisotropic Dirac cones may lead to another classification of Weyl nodes.
      If the tilt is weak with $T({\bf q})<U({\bf q})$ in \eqref{eq4}, the Fermi level $\varepsilon_F$ only cross the upper or lower cone, characterizing an electron or a hole pocket.
      A conventional type-I Weyl semimetal appears , where the Fermi surface shrinks into a point as the Fermi energy crosses the Weyl point \cite{Soluyanov.Gresch.ea:2015:N}.
      If, for a particular direction in reciprocal space, $T({\bf q})$ is dominant over $U({\bf q})$ the tilt becomes large enough to cause a \ac{WP} where electron and hole pockets touch, contrary to the standard case \cite{Soluyanov.Gresch.ea:2015:N}.
      Indeed, for such a direction $\hat{\bf q}$, $T(\hat{\bf q})>U(\hat{\bf q})$ holds.
      Weyl semimetals with such Weyl nodes are called type-II ones.
      Such a situation is claimed to be observable for NbP \cite{Wu.Sun.ea:2017:PRB}.
      Cones at W2 points are indeed strongly tilted in the $k_xk_z$-plane (see Fig.~\ref{fig:3D_fit}).
      The Fermi and tilt velocities in Table~\ref{tab2} give for W2 and NbP a maximum $T(\hat{\bf q})/U(\hat{\bf q})$ ratio for $\hat{\bf q} = (0.420 \ \ 0.000 \ \ 0.908)$. However, with the values from Table I this ratio is still smaller than 1. 
      This is illustrated by three cuts through the two W2 Weyl bands of NbP, calculated \emph{ab initio} in Fig.~\ref{fig:3D_fit} for two views.
      The anisotropy of the upper and lower \ac{3D} Dirac cones are visible in the right panels.
      The left panels clearly show the tilt of these cones.
      However, there is no energetic overlap of the upper and lower cones.
      Consequently, we cannot confirm that NbP is a type-II semimetal.
      This is seemingly in contrast  with the theoretical  findings by \cite{Wu.Sun.ea:2017:PRB}.
      The discrepancy can be explained by noticing that the region were the tilting is large takes place where the band linearization, and therefore the Weyl picture, is not valid anymore.
      Fig. 3c in \cite{Wu.Sun.ea:2017:PRB}) shows indeed  touching points between electrons and holes, which are however away from the linear band regime near the Weyl point, i.e., outside the interpretation of Weyl excitations.
      Consequently, one cannot speak about a type-II Weyl semimetal in the sense of the definition of the linear quantities T and U with T$>$U.
      We are not aware of experiments showing that NbP is a type-II semimetal.
      Transport measurements may give an answer, but  may be influenced by the presence of trivial points, which hinders the analysis of the nodes, and by  carriers at higher or lower energies, where the band linearization, and therefore the Weyl picture, is not valid anymore.
 
    \subsection{\label{sec3d} Effect of tilting on the Density of states}
      The wave functions \eqref{eq5} do not depend on the tilt ${\bf v}_0$ of the Dirac cones.
      Only the energy eigenvalues \eqref{eq4} are modified by the tilt.
      For that reason, matrix elements of the optical transition or the spin current operator should be uninfluenced by the tilt.
      Only the spectral properties may be directly influenced through the change in eigenvalues.
      As an example, the single-particle \ac{DOS}
      \begin{align}\label{eq14}
        D(\varepsilon) = 
          \sum_{j=1,2}
            \sum_{W_j}
              \int
                \frac
                  {d^3{\bf q}}
                  {(2\pi)^2}
                \delta
                \left(
                  \varepsilon - 
                  \varepsilon_\pm({\bf q}) - 
                  \varepsilon_{0j}
                \right)
      \end{align}
      around the Fermi level is investigated.
      Thereby, the energy $\varepsilon_{0j}$ describes the energy displacement of each Weyl node type $W_j$ $(j=1,2)$ with respect to an energy zero.
      In terms of the discussed positions of the band touching points $\Delta\varepsilon_{Fj}$ with respect to the Fermi level, we have $\varepsilon_{0j} \equiv \Delta\varepsilon_{Fj}$.
      Moreover, the contributions of each Weyl node have to be summed up.

      In spherical coordinates one integration can be performed applying the Dirac $\delta$-function
      \begin{align}\label{eq15}
        D(\varepsilon) = 
        \sum_{j=1,2}
          (\varepsilon - \varepsilon_{0j})^2
          \frac{1}{\hbar^3}
          \sum_{W_j}
            \frac{1}{(2\pi)^3}
            \int
              \limits^{2\pi}_{0}
              d\varphi
            \int
              \limits^{+1}_{-1}
              d \xi
              \frac
                {\rm{sgn} ( g_\pm( \varphi, \xi))}
                {g^3_\pm( \varphi, \xi)}
      \end{align}
      with
      \begin{align*}
      \begin{split}
        &g_\pm( \varphi, \xi) = 
          (v_{0x} \cos \varphi + 
            v_{0y} \sin \varphi) \sqrt{1-\xi^2} +
            v_{0z} \xi
        \\
        &\pm
          \left\{ 
            \sum_{j=x,y,z}
              \left[ 
                \left(
                  v_{jx} \cos \varphi + 
                  v_{jy} \sin \varphi
                \right)
                \sqrt{1-\xi^2} + 
                v_{jz} \xi
              \right]^2
          \right\}^{\frac{1}{2}}.
      \end{split}
      \end{align*}
      The \ac{DOS} near a Weyl node ${\bf q}=0$ gives rise to a quadratic energy dependence around the energy of each of the two band touching energies $\varepsilon_{0j}$.
      The cone anisotropy and cone tilt influence the contribution of the two types of Weyl nodes, but in an averaged manner as indicated by the angular integration in \eqref{eq15}.
      In general, the angular integrations can be only carried out numerically.
      In Fig.~SM8  the \emph{ab initio} computed \ac{DOS} of the four Weyl semimetals  displays the contributions of the W1 and W2 Weyl points.
      The difference to the total \ac{DOS} illustrates the effect of the trivial points but also the contributions from other BZ regions.
      Interestingly, the spectra for the phosphides, as well as the arsenides, exhibit major similarities.
      For all monopnictides substantial contributions of the trivial points are visible.

      To explicitly discuss the tilt influence we assume isotropic cones with $\hat{v}_{Fj}={\rm sgn}(det(\hat{v}_F))v_F$, and a tilt along the connection line ${\bf v}_0=(\pm v_0,0,0)$ with the sign as for the left or right node of a pair of type $W_j$.
      Rotating the local coordinate system appropriately
      \begin{align}\label{eq16}
        g_\pm( \varphi, \xi) = 
        \pm 
        v_0 \xi
        \pm 
        v_F,
      \end{align}
      for type-I semimetals we find
      \begin{align}\label{eq17}
        D( \varepsilon) = 
        \sum_{j=1,2}
          N_j
          \frac
            {1}
            {2\pi^2}
          \frac
            {1}
            {\hbar^3v^3_{Fj}}
          \frac
            {1}
            {
              \left[
                1 -
                \left(
                  \frac
                    {v_{0j}}
                    {v_{Fj}}
                \right)^2
              \right]^2
            }
          \left(
            \varepsilon -
            \varepsilon_{0j}
          \right)^2
      \end{align}
      with $N_1=8$ and $N_2=12$.
      In this highly symmetrical case, in contrast to Table~\ref{tab2}, a Weyl node of one type $j=1,2$ is characterized by the three parameters $\varepsilon_{0j}$, $v_{0j}$, and $v_{Fj}$.
      The tilt $v_{0j}$, measured in terms of the variation $v_{Fj}$ of the linear bands, significantly increases the \ac{DOS}, if $|v_{0j}|$ approaches $v_{Fj}$.
      However, approaching the situation of a \ac{WP} in a type-II Weyl semimetal with large tilt the combination of linear approximation for the bands \eqref{eq4} or \eqref{eq16} together with the extension of the wave-vector integration to infinite fails.
      Rather, the integration in \eqref{eq14} has to be limited.

  \section{\label{sec4}Spin Texture in tilted anisotropic Weyl cones}
    \subsection{\label{sec4a}Isolated Weyl node}
\begin{figure*}[h]
    \includegraphics[width=0.95\textwidth,height=0.95\textheight,keepaspectratio]{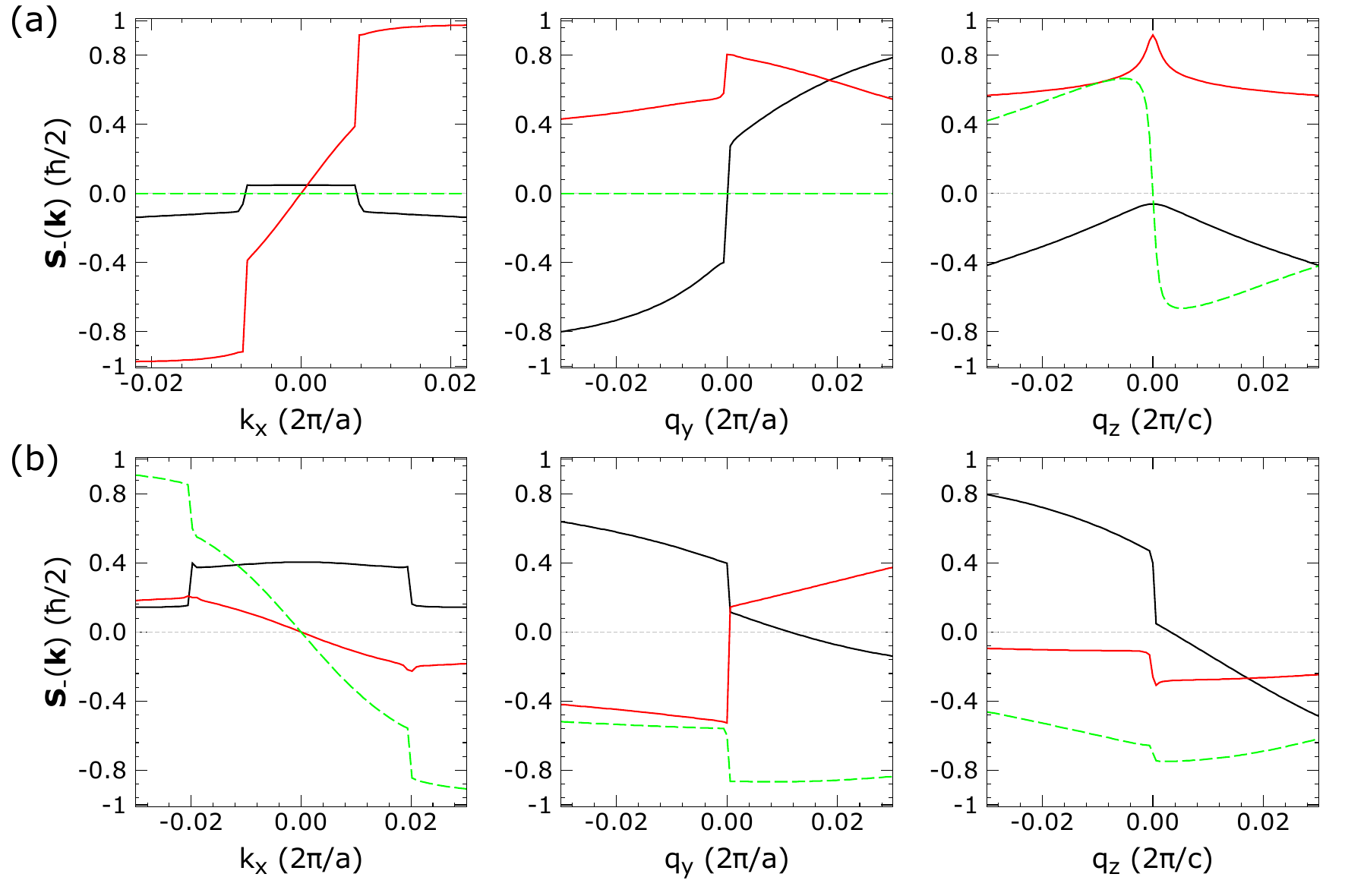}
    \caption{\label{fig:spintxt}
      (Color online) Spin texture ${\bf S}_-({\bf k})$ \eqref{eq18} near a pair of Weyl points at ${\bf k}_W$ and ${\bf k}_W'$ in TaAs calculated \emph{ab initio}.
      Both types W1 (a) and W2 (b) are displayed.
      The three Cartesian components are indicated by solid black ($x$ component), solid red ($y$ component), and dashed green ($z$ component).
    }
  \end{figure*}

      We investigate the expectation value of the spin operator ${\bf s}=\frac{\hbar}{2}\hat{\bm\sigma}$ for Bloch states $|\nu{\bf k}\rangle$,
      \begin{align}\label{eq18}
        {\bf S}_\nu( {\bf k}) = 
        \frac
          {\hbar}
          {2}
        \langle
          \nu{\bf k}
          |
          \hat{\bm\sigma}
          |
          \nu{\bf k}
        \rangle,
      \end{align}
      using the Bloch solutions of the Kohn-Sham equation.
      The results are displayed in Fig.~\ref{fig:spintxt}.
      Thereby, we restrict ourselves to the lower Weyl band and wave vectors ${\bf k}$ around a Weyl point ${\bf k}_W$.
      In terms of the corresponding eigenstates \eqref{eq5} of the model Hamiltonian \eqref{eq2} the spin expectation value of the lower Weyl band is
      \begin{align}\label{eq19}
        {\bf S}_- ( {\bf q}) = 
        \frac
          {\hbar}
          {2}
        \langle
          \Psi_{-{\bf q}}
          |
          \hat{\bm\sigma}
          |
          \Psi_{-{\bf q}}
        \rangle.
      \end{align}
      Using the expressions \eqref{eq5}, \eqref{eq6}, and \eqref{eq7}, one obtains
      \begin{align}\label{eq20}
        {\bf S}_- ( {\bf q}) = 
        -\frac
          {\hbar}
          {2}
        \frac
          {1}
          {\sqrt{
            \sum\limits_{j=x,y,z}({\bf v}_j{\bf q})^2
            }
          }
        \left(
          {\bf v}_x{\bf q},
          {\bf v}_y{\bf q},
          {\bf v}_z{\bf q}
        \right)
        =
        -\frac
          {\hbar}
          {2}
        \frac
          {\hat{v}_F{\bf q}}
          {|\hat{v}_F{\bf q}|}
      \end{align}
      with the velocity vectors ${\bf v}_j$ ($j=x,y,z$) defined by columns or rows of the matrix of the Fermi velocities \eqref{eq3}.

      The model result \eqref{eq20} is very interesting: (i) The spin texture of the lowest Weyl band in the surroundings of a Weyl node does not depend on the tilt of the cones or the kinetic energy $T({\bf q})$ \eqref{eq4}, rather only on the shape of the cones.
      (ii) The magnitude of the spin vector \eqref{eq18}, $|{\bf S}({\bf q})|=\frac{\hbar}{2}$, is conserved and independent of the orientation of the wave vector ${\bf q}$.
      (iii) For complete isotropic cones with  Fermi velocities, $v_{ij}={\rm sgn}(det(\hat{v}_F))v_F\delta_{ij}$, the spin texture is ${\bf S}({\bf q})=-\frac{\hbar}{2}{\rm sgn}(det(\hat{v}_F))\frac{{\bf q}}{|{\bf q}|}$.
      This is the well-known fact that the spin of an isotropic Weyl fermion is parallel or antiparallel to its wave vector \cite{Jia.Xu.ea:2016:NM} and the spin texture forms a 'hedgehog' around a Weyl node\cite{Burkov:2016:NM}.
      However, in a real Weyl semimetal the Weyl fermions occur near a \ac{WP} at  ${\bf k}_W$  antiparallel to the wave vector, ${\bf q}={\bf k}-{\bf k}_W$, rotated by the matrix $\hat{v}_F$ \eqref{eq3} of the Fermi velocities.
      This is an obvious generalization of the Weyl fermion picture in \ac{3D} Weyl semimetals due to the anisotropy of the Dirac cones.
      Nevertheless, the picture of strong correlation of spin and wave-vector orientation still helps to interpret transport measurements \cite{Zhang.Tong.ea:2016:PRB}.

      The analytical prediction \eqref{eq20} is compared with results of \emph{ab initio} calculations in Fig.~\ref{fig:spintxt} for the lower Weyl band and ${\bf k}$ near ${\bf k}_W$ and ${\bf k}_W^\prime$.
      In Fig.~\ref{fig:spintxt} we display the spin expectation values \eqref{eq18} versus the wave vector for TaAs with the largest distance between two Weyl points in one pair.
      Actually, we have chosen the W1 pair at ${\bf k}_W=(k_{Wx}, k_{Wy}, k_{Wz})$ and ${\bf k}_W^\prime=(-k_{Wx}, k_{Wy}, k_{Wz})$ with $k_{Wx} = 0.0078 \frac{2\pi}{a}$, $k_{Wy} = 0.5103 \frac{2\pi}{a}$ and $k_{Wz}=0$ and the W2 pair at ${\bf k}_W=(k_{Wx},k_{Wy},k_{Wz})$ with and ${\bf k}_W'=(-k_{Wx},k_{Wy},k_{Wz})$ with $k_{Wx}=-0.0198\frac{2\pi}{a}$, $k_{Wy}=0.2818\frac{2\pi}{a}$, and $k_{Wz}=0.5905\frac{2\pi}{c}$, where the nodes are separated along the $k_x$-axis.

      The comparison demonstrates some similarities but also contradictions between the \emph{ab initio} calculations and the model prediction \eqref{eq20}.
      The general behavior of the classical Weyl fermion picture is modified due to the wave-vector rotation by the matrix of the Fermi velocities as shown in expression \eqref{eq20}.
      Hence the components of the spin vector are no longer parallel to the momentum around a \ac{WP}.
      This especially holds for its orientation and accompanied change in sign.
      Far away from the Weyl nodes, for a Cartesian component $j$ along the same component of the wave vector, it holds $S_j(q_j)=\pm\frac{\hbar}{2}$.
      The finite second Cartesian component $S_j$ perpendicular of the wave-vector direction is related to the finite off-diagonal Fermi velocities in Table~\ref{tab2}.
      Deviations from this simple picture of anisotropic Weyl fermions along the $q_z$-direction are due to the fact that all Fermi velocities $v_{jz}$($j=x,y,z$) in Table~\ref{tab2} are of the same order of magnitude, in contrast to the components $v_{jx}$ $(v_{jy})$ of the vectors ${\bf v}_x$ and ${\bf v}_y$, respectively, where the diagonal elements are the largest ones.

      The spin texture calculated \emph{ab initio} for the lower Weyl band $\nu = -$ in Fig.~\ref{fig:spintxt} is strongly influenced by the facts that the spin is a non-conserving quantity due to the \ac{SOC} and the Weyl nodes appear in pairs with such a small distance that the picture of isolated Weyl nodes is hardly valid.
      The Cartesian components $S_x(k_x)$ and $S_y(k_x)$ along the connection line of the pair points exhibit a similar behavior independent of W1 or W2.
      $S_x(k_x)$ is almost constant but changing sign crossing a Weyl point.
      $S_y(k_x)$ varies linearly with a zero value exactly between the two Weyl points.
      The different signs are a consequence of the interchanged chirality of the Weyl point W1 and W2 at $k_{Wx}$.
      $S_z(k_x)$ shows qualitative differences between W1 and W2 pairs.
      While $S_z(k_x) = 0$ in the W1 case, it shows a similar behavior as $S_y(k_x)$ for W2.
      The different behavior of the components of the spin expectation value vector along the pair connection line may be a consequence of the varying distance of the Weyl nodes in a W2 pair with the Weyl semimetal (see Table~SM1) and the 2D band character for W1 (see Fig.~SM7).
      Differences between W1 and W2 are visible in Fig.~\ref{fig:spintxt} also for the spin component along the two Cartesian directions $q_y$ and $q_z$.
      Different signs may be a consequence of the different chiralities ${\rm sgn}(det \ \hat{v_F}$) (see Table~\ref{tab2}) within the chosen geometries.
      However, contrary to the model expectation, sign changes crossing the Weyl node only happen for $S_x(q_y)$, $S_z(q_z)$ in the W1, and $S_y(q_y)$, $S_x(q_z)$ in the W2 case.
      The picture of "hedgehogs" with opposite orientation of the spin expectation values, derived for isotropic and untilted pair of Weyl nodes, is not valid for a real Weyl semimetal since a similar behavior is found for the other three materials TaP, NbAs, and NbP (not displayed in Fig.~\ref{fig:spintxt}).

    \subsection{\label{sec4b}Effect of pairing}
      Strong deviations from the picture of anisotropic Weyl fermions illustrated by formula \eqref{eq20}, appear in Fig.~\ref{fig:spintxt} (most left panels) for both types of Weyl nodes for wave vectors $k_x$ along the connection line of two \acp{WP} forming a pair.
      This holds especially for wave vectors $|k_x|<k_{Wx}$, i.e., $q_x<0$ (right node) and $q_x>0$ (left node in a pair).
      In addition, as a consequence of the chirality change, the component $S_x(k_x)$ changes the sign, and it holds $S_x(\pm k_{Wx}) = 0$.
      This observation cannot be explained in the approximation of an isolated Weyl point, i.e., the Hamiltonian $\hat{H}_W$ \eqref{eq2}.

      For the purpose of interpretation of the findings in the left panels of Fig.~\ref{fig:spintxt}, we apply the minimal two-node model for a \ac{WP} pair of a Weyl semimetal \cite{Lu.Zhang.ea:2015:PRB,Zhang.Lu.ea:2016:NJoP,Lu.Shen:2017:FoP} described by Hamiltonian $\hat{H}_{2W}$ \eqref{eq8}. It focuses on the anisotropy due to Weyl points occurring in pairs but omits the details of the band structure near points  ${\bf k}_W$ and ${\bf k^{\prime}}_W$, which are not located at high-symmetry lines.

      With the eigenvectors \eqref{eq5}, \eqref{eq6}, and \eqref{eq7} but with the modification \eqref{eq10} one obtains the model spin texture
      \begin{align}\label{eq21}
      \begin{split}
          {\bf S}_- ({\bf Q}) =
            &\frac
              {\hbar}
              {2}
            \frac
              {1}
              {
              \left[
                \left(
                  \frac
                    {v_{xx}}
                    {2k_{Wx}}
                \right)^2 
                \left(
                  k^2_{Wx} -
                  Q^2
                \right)^2
                + 
                v^2_{yy}(q^2_y+q^2_z)
              \right]^{\frac{1}{2}}
              } \\ 
            & \times 
            \left(
              \frac
                {v_{xx}}
                {2k_{Wx}}
              (k^2_{Wx}-Q^2),
              v_{yy}q_y,
              v_{yy}q_z
            \right)
            .
          \end{split}
      \end{align}
      Close to the two Weyl points $(\pm k_{Wx},k_{Wy},k_{Wz})$ one finds still deformed 'hedgehogs' with
      \begin{align}\label{eq22}
        {\bf S}_- ( {\bf Q}) = 
        \mp
        \frac
          {\hbar}
          {2}
        \frac
          {1}
          {
          \left[
            v^2_{xx}q_x^2 + 
            v^2_{yy} (q_y^2 + q^2_z)
          \right]^{\frac{1}{2}}}
        \left(
          v_{xx}q_x,
          v_{yy}q_y,
          v_{yy}q_z
        \right)
        ,
      \end{align}
      where the wave vector $q_x=k_x-k_{Wx}$ ($q_x=k_x+k_{Wx}$) near the right (left) Weyl point has been introduced.
      The spin textures point in opposite directions at the two \acp{WP} in the direction of the rotated reduced wave vector.
      More in general, the findings \eqref{eq21} for the spin texture exhibit significant modifications of the $x$-component of the spin texture along the connection line of the two Weyl points.
      Especially, for vanishing perpendicular components $q_y=q_z=0$ the component $S_x(k_x,0,0)$ shows a zero at $k_x=k_{Wx}$ and changes its sign, according to $sgn(k_{Wx}^2 - k_{x}^2)$, but remain constant.
      This fact is in good agreement with the \emph{ab initio} calculated results in Fig.~\ref{fig:spintxt} (left panels).
      However, also for wave-vector variations along $y$- or $z$-directions, e.g. the component $S_x(0,q_y,0)\sim(k^2_{Wx}-q^2_y)$, it is finite but with varying absolute values and signs.
      Qualitatively such a behavior is observed in Fig.~\ref{fig:spintxt} also in the spin texture computed \emph{ab initio} for many components and directions.
      However there are also deviations which cannot be described by the simple model results \eqref{eq21}.
      
  \section{\label{sec5}Topological properties}
    \subsection{\label{sec5a}Berry connection and curvature in anisotropic cones}
      Geometrical phases that characterize the topological properties of Bloch bands play a fundamental role in the band theory of solids \cite{Shen:2014:Book}.
      For a Bloch band $\nu$ with the Bloch factor $U_{\nu{\bf k}}$ its Berry connection or Berry vector potential is defined as \cite{Berry:1984:PRSLAMPS}
      \begin{align}\label{eq23}
        {\bf A}^\nu( {\bf k}) = 
        i
        \langle
          U_{\nu{\bf k}}
          |
          {\bm\nabla}_{\bf k}
          |
          U_{\nu{\bf k}}
        \rangle.
      \end{align}
      The Berry connection corresponds to a gauge field defined in the ${\bf k}$-space, similar to the vector potential of electromagnetic fields in real space \cite{Ando:2013:JPSJ}.
      Its rotation
      \begin{align}\label{eq24}
        {\bm\Omega}^\nu( {\bf k}) = 
        {\bm\nabla}_{\bf k}
          \times
        {\bf A}^\nu({\bf k}) = 
        {\rm curl}{\bf A}^\nu
        \left(
          {\bf k}
        \right),
      \end{align}
      known as Berry curvature, is better defined.
      It is an intrinsic property of the band structure, because it depends only on the Bloch wave function \cite{Xiao.Chang.ea:2010:RMP}.

\begin{figure*}[h]
    \includegraphics[width=0.95\textwidth,height=0.95\textheight,keepaspectratio]{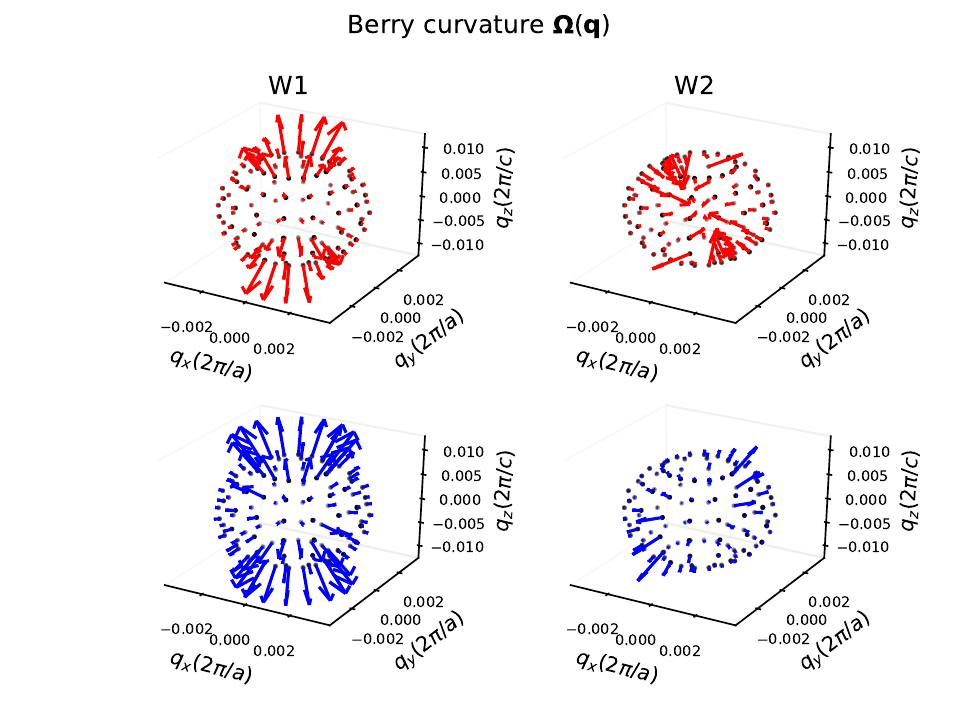}
    \caption{\label{fig:berry}
      (Color online) Berry curvature calculated analytically using \eqref{eq25} (red) and \eqref{eq26} (blue) near a W1 (left panels) or W2 (right panels) node for TaAs.
    }
  \end{figure*}

      Here, we are mainly interested in the topological properties near a Weyl node ${\bf k}_W$, i.e., for ${\bf q}={\bf k}-{\bf k}_W$, of the lower Weyl band $\nu=-$.
      The Berry curvature of this band is illustrated in Fig.~\ref{fig:berry} for a W1 and a W2 Weyl point in TaAs.
      Very close to the Weyl point its wave-vector dependence can be illustrated by the model \eqref{eq2}.
      With the Bloch factors \eqref{eq5} with $f({\bf q})$ and $\theta({\bf q})$ given in \eqref{eq7}, one finds for the lower Weyl band $\nu = -$
      \begin{align}\label{eq25}
        \begin{split}
          {\bf A}( {\bf q}) &=
          -\frac
            {1}
            {2}
          f
          \left(
            {\bf q}
          \right)
          {\bm \nabla}_{\bf q}
          \theta
          \left(
            {\bf q}
          \right)
          ,
          \\
          {\bm\Omega}( {\bf q}) &=
          -\frac
            {1}
            {2}
          \left(
            {\bm\nabla}_{\bf q} f({\bf q})
          \right)
            \times
          \left(
            {\bm\nabla}_{\bf q}\theta({\bf q})
          \right).
        \end{split}
      \end{align}
      It remains uninfluenced by the tilt.
      The quantities for the upper Weyl band are just the negative ones.
      The explicit expressions using the full Fermi-velocity matrix \eqref{eq3} are rather complex.
      In the high-symmetry case, with a diagonal matrix $v_{ij} = v_{ii} \delta_{ij}$ one finds
      \begin{align}\label{eq26}
        \begin{split}
          {\bf A}( {\bf q}) &=
          \frac
            {1}
            {2}
          f({\bf q})
          \frac
            {v_{xx}v_{yy}}
            {(v_{xx}q_x)^2+(v_{yy}q_y)^2}
          \left(
            q_{y},
            -q_x,
            0
          \right),
          \\
          {\bf \Omega}( {\bf q}) &= 
          \frac
            {1}
            {2}
          \frac
            {v_{xx}v_{yy}v_{zz}}
            {
            \left[
              (v_{xx}q_x)^2 +
              (v_{yy}q_y)^2 +
              (v_{zz}q_z)^2
            \right]^{\frac{3}{2}}
            }
          {\bf q}.
        \end{split}
      \end{align}
      In this limit the Berry curvature points into the direction of the wave vector.
      In the fully isotropic limit $v_{xx} = v_{yy} = v_{zz} = {\rm sgn}(det(\hat{v}_F))v_F$ the vector \eqref{eq26} defines a flux through any surface in ${\bf q}$-space, enclosing the Weyl node.
      Indeed it forms a "hedgehog" \cite{Burkov:2016:NM}.

      The gauge-invariant Berry phase $\gamma^\nu_{\rm Berry}$ of a Bloch band $\nu$ is defined as a surface integral of the Berry curvature, where an arbitrary surface enclosed by the path $\mathcal{C}$ in the parameter space is used.
      With the Stokes theorem this definition can be rewritten as a path integral in the parameter space \cite{Xiao.Chang.ea:2010:RMP}, $\gamma^\nu_{\rm Berry}=\oint_{\mathcal{C}}d{\bf k}{\bf A}^\nu({\bf k})$, where the wave vector ${\bf k}$ is adiabatically rotated anticlockwise along the path $\mathcal{C}$.
      In the model case \eqref{eq2}, the Berry phase can be easily calculated studying a small surrounding of a WP.
      Choosing a small circle in the $q_xq_y$-plane in a large distance $q_z$ from the Weyl point at ${\bf q}=0$, the approximation \eqref{eq26} yields $\gamma_{\rm Berry}={\rm sgn}(det \ \hat{v}_F)\pi$ for the lower Weyl band.
      Outside the Weyl point this band is, therefore, non-trivial, i.e., topological.
      \cite{Klotz.Wu.ea:2016:PRB}
      The experimental determination of the $\pi$ Berry phase has been performed under quantized conditions of an external magnetic field for TaP and NbP \cite{Hu.Liu.ea:2016:SR,Sergelius.Gooth.ea:2016:SR}.
      
      For more clarity, we display in Fig.~\ref{fig:berry} the Berry curvature of the lower Weyl band near a WP only using model results for TaAs parameters .
      In the W1 case, the 'hedgehog' character is nearly visible.
      This is a consequence of the same signs of the Fermi velocities in Table~\ref{tab2} and the fact that the matrix $\hat{v}_F$ is nearly diagonal.
      The indicated small anisotropy is mainly due to the tetragonal symmetry.
      In the W2 case, stronger deviations from the 'hedgehog' behavior are observable, independently of using the more complex \eqref{eq25} or the simplified \eqref{eq26} models.
      This finding suggests that the Weyl behavior of the low-energy electronic excitations near a W2 Weyl node is more complex.
      The trivial (fully isotropic) Weyl picture has to be replaced by a more complicated behavior of the Berry curvature, even in the model case \eqref{eq2}.
      While the more reliable description \eqref{eq25} of a W2 point shows for many {\bf q}-points a behavior in agreement with $det(\hat{v}_F)<0$ (see Table~\ref{tab2}), the neglect of the off-diagonal elements of $\hat{v}_F$ gives rise to a behavior in agreement with the approximation $det(\hat{v}_F)>0$ used in \eqref{eq26}.

    \subsection{\label{sec5b}Monopole charge and chirality}
      The Berry curvature around a \ac{WP} allows to define formally a magnetic monopole charge by the Berry curvature flux threading a closed surface $F$ that encloses the origin ${\bf q}=0$,
      \begin{align}\label{eq27}
        c = 
        \frac
          {1}
          {2\pi}
        \oiint
          \limits_F
          d{\bf F}
          ({\bf q})
          {\bm\Omega}({\bf q})
        .
      \end{align}
      Expression \eqref{eq26} indicates that using an appropriately deformed ellipsoid $F$ in \eqref{eq27} leads to
      \begin{align}\label{eq28}
        c = 
        {\rm sgn}
        \left(
          det(\hat{v}_F)
        \right)
        .
      \end{align}
      The sign of the 
       (magnetic) monopole in reciprocal space is related to its chirality ${\rm sgn}(det \ \hat{v}_{F})$.

      For the other Weyl point in the pair, because of $v_{xx} \rightarrow -v_{xx}$, the opposite sign results from \eqref{eq28}.
      Consequently, it possesses the opposite chirality.
      Thus, the total monopole charge of a pair of Weyl points is zero, in agreement with the fact that the total magnetic charge in a band structure must be zero \cite{Nielsen.Ninomiya:1981:NPB,Nielsen.Ninomiya:1981:NPBa}.
      The opposite signs of the monopole charges describe that the two \acp{WP} in one pair act as a source or sink, respectively, of the Berry curvature.

    \subsection{\label{sec5c}Influence of pairing}
      The results \eqref{eq25} can be generalized to the case of the two-node model \eqref{eq8}.
      With $\theta({\bf Q})$ and $f({\bf Q})$ from \eqref{eq10}, one obtains
      \begin{align}\label{eq29}
        \begin{split}
          {\bm\Omega}( {\bf Q}) =
          &\frac
            {
              v^2_{yy}
              \left(
                \frac
                  {v_{xx}}
                  {2k_{Wx}}
              \right)
            }
            {
              \left[
                \left(
                  \frac
                    {v_{xx}}
                    {2k_{Wx}}
                \right)^2
                (k^2_{Wx} - Q^2)^2 +
                v^2_{yy} (q^2_y+q^2_z)
              \right]^\frac{3}{2}}
          \\
          &\times
          \left(
            -\frac
              {1}
              {2}
            (k^2_{Wx}-Q^2) - 
            q^2_y - 
            q^2_z,
            k_xq_y,
            k_xq_z
          \right)
          .
        \end{split}
      \end{align}
      Its linearization in $q_x$ near one of the two Weyl points in the pair with $k_x=\pm k_{Wx}+q_x$ leads to the result \eqref{eq26} for $v_{yy}=v_{zz}$ but with opposite signs thereby indicating opposite signs of the monopole charge and the chirality of the two \acp{WP} in the pair.

      The Berry curvature \eqref{eq29} allows to define a Zak phase \cite{Zak:1989:PRL} instead of a Berry phase of the lower Weyl band.
      It is the Berry phase integrated along a path $\mathcal{C}$ in one plane of the momentum space traversing the connection line between the two \acp{WP} of one pair \cite{Kim.Lee.ea:2016:NC}.
      Zak phases have been measured for Bloch bands of one-dimensional photonic lattices \cite{Atala.Aidelsburger.ea:2013:NP}.
      Applying Stokes theorem and the two-node model \eqref{eq29} it holds
      \begin{align}\label{eq30}
        \gamma_{\rm Zak}( k_x) = 
        \iint
          \limits_{ q_y q_z -{\rm plane}}
          d{\bf F} ({\bf Q})
          {\bm\Omega}({\bf Q}).
      \end{align}
      One finds for an anticlockwise path in this plane
      \begin{align}\label{eq31}
        \gamma_{\rm Zak}(k_x)=2\pi{\rm sgn}(v_{xx})\theta(k_{Wx}-|k_x|).
      \end{align}
      The corresponding winding or Chern number is
      \begin{align}\label{eq32}
        C_{\rm Chern}( k_x) =
        {\rm sgn} (v_{xx})
        \theta(k_{Wx}-|k_x|).
      \end{align}
      The net Berry phase accumulated in a $q_yq_z$-plane between a pair of \acp{WP} along $k_x$ induces a nonzero Chern number $C_{\rm Chern}(k_x)={\rm sgn}(v_{xx})$, while its value is zero, $C_{\rm Chern}(k_x)=0$, for $k_x < -k_{Wx}$ or $k_x > k_{Wx}$.
      That means that each \ac{2D} slice in momentum space, which does not contain any Weyl nodes, can be thought of as a Chern insulator.
      To be topological for $|k_x|<k_{Wx}$ means that at the boundaries of such planes in real space topological edge states may occur with wave vectors in the surface \ac{BZ} in the $k_xq_y$-plane, if the surface normal is  chosen parallel $q_z$.
      Indeed, such states have been observed by \ac{ARPES} as so-called Fermi arc states, mainly for anion-terminated, as-cleaved surfaces of TaAs \cite{Xu.Belopolski.ea:2015:S,Lv.Weng.ea:2015:PRX,Yang.Liu.ea:2015:NP}, TaP \cite{Liu.Yang.ea:2015:NM}, and NbP \cite{Liu.Yang.ea:2015:NM,Belopolski.Xu.ea:2016:PRL}.
      The small values of $k_{Wx}$ (see Table~SM1), together with the condition $k_{Wx}>|k_x|$ in \eqref{eq31} and \eqref{eq32} indicate that low-energy Weyl fermions near the Fermi energy of bulk NbP are difficult to observe \cite{Shekhar.Nayak.ea:2015:NP,Zhang.Tong.ea:2016:PRB}.

  \section{\label{sec6}Summary and Conclusions}
    By means of \emph{ab initio} band structure calculations for Weyl semimetals and model studies for isolated or paired Weyl nodes we have investigated the influence of tilt, anisotropy and pairing of these nodes.
    As examples the \ac{bct} Weyl semimetals TaAs, TaP, NbAs, and NbP have been chosen.
    The \emph{ab initio} calculations allowed us to determine the \ac{BZ} position of each \ac{WP}, the tilt of its Weyl cones, and the full symmetric matrix of the Fermi velocities, which however clearly describes anisotropy, deformation and particle-antiparticle inequivalence of the cones near a \ac{WP}.
    Together with the computed Fermi levels the possibility of low-energy Weyl fermion excitations and their density of states have been analyzed.
    The discussion of the pairs of Weyl nodes is completed by that of trivial points, which give rise to hole pockets but are not protected by symmetry.
    Therefore, their number varies with the different cations and anions.

    The one- and two-node models for the electronic structure of a Weyl pair, in combination with the computed explicit node parameters, have been applied to the understanding of the total \ac{DOS}, the spin texture, and the Berry  curvature.
    The single-particle \ac{DOS} significantly depends on the tilting and the positions of the W1 and W2 Weyl nodes in energy space, but also on the trivial points.
    The spin texture is significantly influenced by the anisotropy of the Dirac cones.
    The fact that wave vector and spin are parallel or antiparallel, known from the isotropic Weyl fermion picture, is violated.
    The spin around a \ac{WP} is rotated by the matrix of the Fermi velocities.
    The node pairing destroys the conventional picture, in particular for wave vectors near the midpoint of the two nodes, where spin components can change their signs.
    The cone anisotropy also strongly influences the Berry curvature around a \ac{WP}.
    However, it still defines a monopole charge of such a node.
    The pairing of nodes is responsible for non-vanishing Zak phases and Chern numbers of planes perpendicular  to pair connection line, indicating the topological character for such wave vectors.

\vskip 1cm
  \begin{acknowledgments}
    F.B. acknowledges travel support by INFN Tor Vergata.
    O.P. acknowledges financial fundings from the EU project CoExAN (GA644076).
    CPU time was granted by CINECA HPC center.
  \end{acknowledgments}

  \bibliography{bibliography}

  \newpage




  \newpage

  \begin{table}[h!]
    \caption{
      Fits of $\hat{v}_{F}$ and $\bm{v_0}$ for Weyl nodes on the $k_x>0$ side of the mirror plane.
      All velocities are in units of $10^5 \ {\rm m} \times {\rm s}^{-1}$
      }
    \begin{tabular}{ | c | c c c c | c c c c |}  \hline
                        & \multicolumn{4}{c|}{W1+}          & \multicolumn{4}{c|}{W2+}           \\
                        & TaAs   & TaP    & NbAs   & NbP    & TaAs   & TaP    & NbAs   & NbP     \\ \hline
      $v_{xx}$          & 3.963  & 4.354  & 3.699  & 4.600  & 3.220  & 0.273  & 2.826  & 2.696   \\
      $v_{yy}$          & 2.318  & 2.592  & 1.684  & 2.277  & 0.291  & 1.783  & 1.611  & 1.629   \\
      $v_{zz}$          & 0.212  & 0.233  & 0.421  & 0.079  & 1.659  & 1.505  & 2.455  & 2.378   \\
      $v_{xy}$          & 0.393  & 0.405  & 0.030  & 0.084  & 1.127  & 1.867  & 0.263  & 0.221   \\
      $v_{xz}$          & 0.000  & 0.000  & 0.000  & 0.000  & 0.661  & 2.599  & 0.248  & 0.210   \\
      $v_{yz}$          & 0.000  & 0.000  & 0.000  & 0.000  & 2.464  & 0.311  & 1.164  & 1.201   \\ \hline
      $det(\hat{v}_F)$  & 1.92   & 2.60   & 2.62   & 0.83   & -16.56 & -13.57 & 7.23   & 6.48    \\ \hline
      $v_{0x}$          & -1.603 & -1.528 & -1.599 & -1.228 & -0.989 & -1.038 & -0.521 & -0.665  \\
      $v_{0y}$          & 1.004  & 1.242  & -0.024 & 0.683  & 0.944  & 0.601  & 0.457  & 0.023   \\
      $v_{0z}$          & 0.000  & 0.000  & 0.000  & 0.000  & 1.409  & 1.543  & 1.223  & 1.105   \\ \hline
    \end{tabular}
    \label{tab2}
  \end{table}

\end{document}


\title{Influence of anisotropy, tilt and pairing of Weyl nodes: \\The Weyl semimetals TaAs, TaP, NbAs, and NbP}

  \author{Davide Grassano and Olivia Pulci and Elena Cannuccia}
  \affiliation{%
    Dipartimento di Fisica, Universit\`a "Tor Vergata" Roma and INFN, via della Ricerca Scientifica, I-00133 Rome, Italy
    }%

  \author{Friedhelm Bechstedt}
  \affiliation{%
    Institut f\"ur Festk\"orpertheorie und -optik, Friedrich-Schiller-Universit\"at Jena, Max-Wien-Platz 1, 07743 Jena, Germany
    }%

  \date{\today}
  \maketitle

  \renewcommand{\theequation}{SM\arabic{equation}}
  \renewcommand{\thefigure}{SM\arabic{figure}}
  \renewcommand{\thetable}{SM\arabic{table}}
  \renewcommand{\bibnumfmt}[1]{[S#1]}
  \renewcommand{\citenumfont}[1]{S#1}

  \section{\label{sec1} Weyl vs Trivial points}

\begin{figure}[h]
      \includegraphics[scale=0.70]{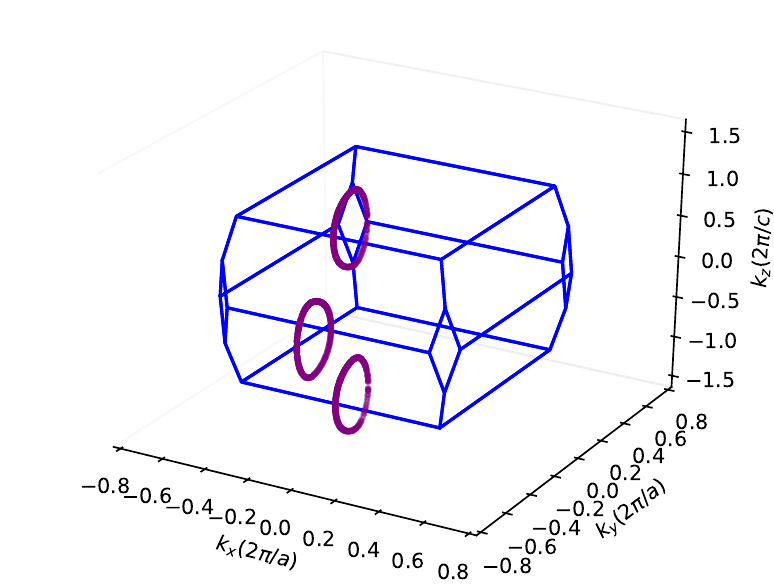}
      \caption{
        The nodal lines for the $k_x = 0$ and $k_y < 0$ quadrant are indicated as a stream of purple dots.
        }\label{fig:SM3}
    \end{figure}

\begin{figure}[h]
      \includegraphics[scale=0.70]{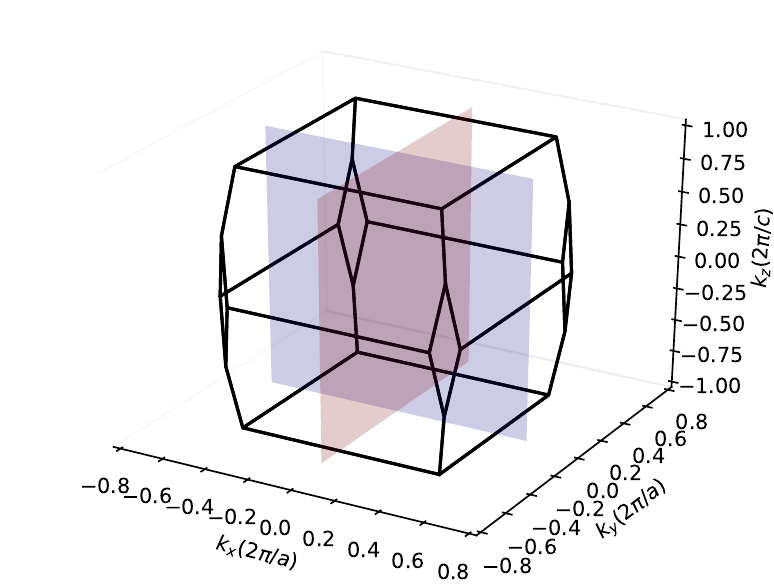}
      \caption{
        The mirror planes $M_x$($M_y$) are indicated in red(blue).
        }\label{fig:SM4}
    \end{figure}

\begin{figure}[h]
      \includegraphics[scale=0.70]{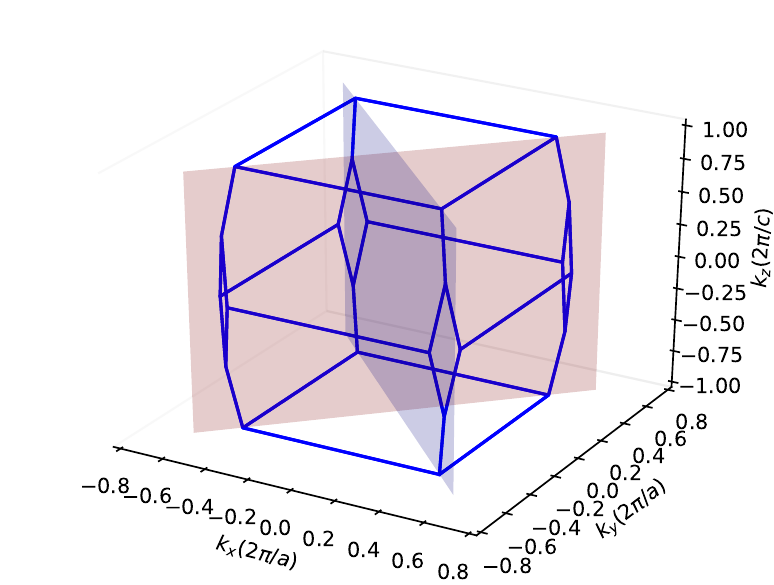}
      \caption{
        The mirror planes $M_{xy}$($M_{-xy}$) are indicated in red(blue).
        }\label{fig:SM5}
    \end{figure}

 The crystals  of TaAs, TaP, NbAs, NbP contain two mirror planes, $M_x$ and $M_y$ and two glide mirror planes, $M_{xy}$ and $M_{-xy}$ (Fig.\ref{fig:SM4},\ref{fig:SM5}).
      Thereby, the Cartesian axes $x$ and $y$ are perpendicular to the tetragonal axis, which defines the $z$ axis.
      The mirror symmetries protect four nodal rings (Fig.\ref{fig:SM3}
      in the mirror planes, which are represented by gapless points when \ac{SOC} is not included.
      With \ac{SOC} a single nodal ring evolves into three pairs of (gapless) band touching points, one pair is in the $k_z=0$ plane of the \ac{BZ} (labeled W1) and two pairs off the $k_z=0$ plane (labeled W2).
      
       For one WP at ${\bf k}_W$, the time-reversal symmetry requires another one at $-{\bf k}_W$.
      Because of their pairing, there must exist two more Weyl points at ${\bf k}'_W$ and $-{\bf k}_W'$.
      Pairs of \acp{WP} are formed at ${\bf k}_W$ and ${\bf k}_W'$ as well as $-{\bf k}_W$ and $-{\bf k}_W'$.
      In the W1 case the other four \acp{WP} follow because of the $M_x$ and $M_y$ mirror operations.
      In the W2 case the total number of ${\bf k}_W$ points is doubled because of their occurrence in planes at $k_z=k_{Wz}$ and $k_z=-k_{Wz}$.
      The two \acp{WP} in one pair are close to each other in $k_x$- or $k_y$-direction around $k_x=0$ and $k_y=0$, respectively.
      The ${\bf k}$-space splittings of such a pair along these two Cartesian directions amount to 0.0156 (TaAs), 0.0154 (TaP), 0.0052 (NbAs), and 0.0058$\frac{2\pi}{a}$ (NbP) in the W1 case or 0.0396 (TaAs), 0.0322 (TaP), 0.0128 (NbAs), and 0.0094$\frac{2\pi}{a}$ (NbP) in the W2 one.
      The positions given in Table~\ref{SM:tab1} fully obey the discussed $M_x$- and $M_y$-mirror symmetries up to four digits.
      This fact underlines the accuracy of our numerical procedures to determine the Weyl node positions.

      For TaAs the absolute positions of W1 and W2 are in very good agreement with similar \ac{DFT} calculations \cite{Huang.Xu.ea:2015:NC,Xu.Belopolski.ea:2015:S,Lee.Xu.ea:2015:PRB,Buckeridge.Jevdokimovs.ea:2016:PRB}.
      This also holds for the calculations for the other three monopnictides, if the same denotations of W1 and W2 as well as the same $k_z$-zero are applied \cite{Lee.Xu.ea:2015:PRB,Xu.Autes.ea:2017:PRL}.
      Also the measured ${\bf k}$-space separations between two nodes of a pair, e.g. for W2 in TaAs \cite{Xu.Belopolski.ea:2015:S,Souma.Wang.ea:2016:PRB}, or the absolute position of the \acp{WP} in NbP \cite{Xu.Autes.ea:2017:PRL} agree well with the computed results.
  \begin{figure}[h]
    \includegraphics[trim={1cm 0cm 0cm 0cm},clip,width=0.5\textwidth,height=0.95\textheight,keepaspectratio]{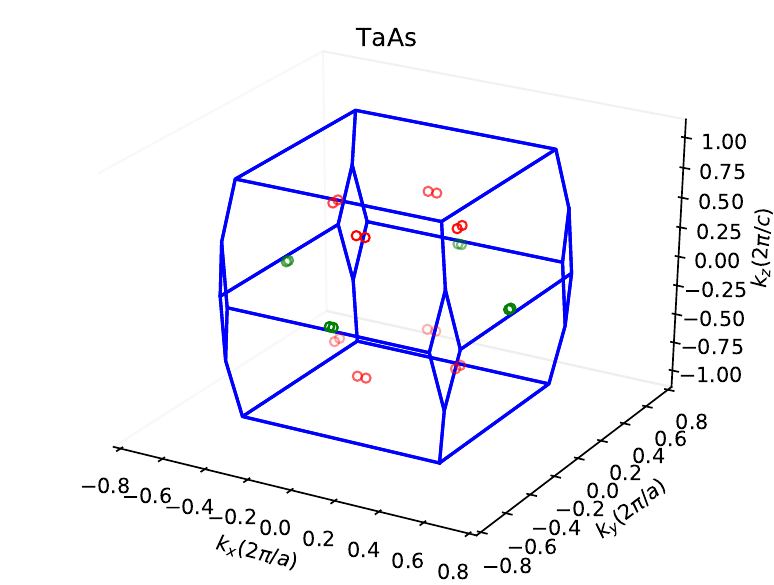}
    \caption{
      (Color online) Distribution of the 12 pairs of Weyl points (green circles: W1, red circles: W2) over the BZ (blue lines) for TaAs.
      }\label{fig:SM1}
  \end{figure}

    \begin{figure}[h]
      \includegraphics[scale=0.70]{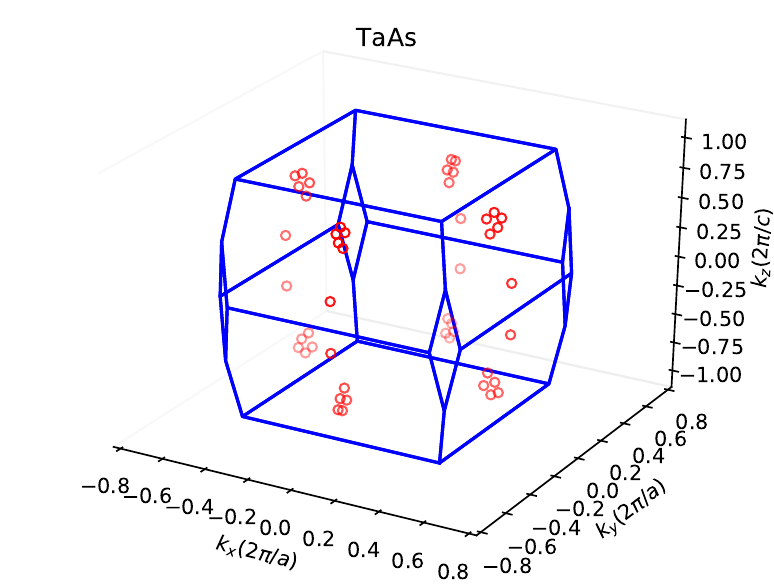}
      \caption{
        Dispersion of the trivial points (red circles) within the BZ for TaAs.
        For clarity, no symmetry operation has been applied to reduce the number of equivalent points.
        }\label{fig:SM2}
    \end{figure}
    \FloatBarrier

  \section{Nodal lines}
    The nodal lines can be observed by running the calculation without SOC.
    Each point of the nodal line is selected, from a band structure calculation over a fine mesh of k-points on the Y=0 plane, using the condition $E_{gap} < 1 meV$
    \FloatBarrier

  \section{Symmetry planes}
    \FloatBarrier

    \FloatBarrier

\section{Electronic properties}
  \begin{figure*}[h]
    \includegraphics[width=0.95\textwidth,height=0.95\textheight,keepaspectratio]{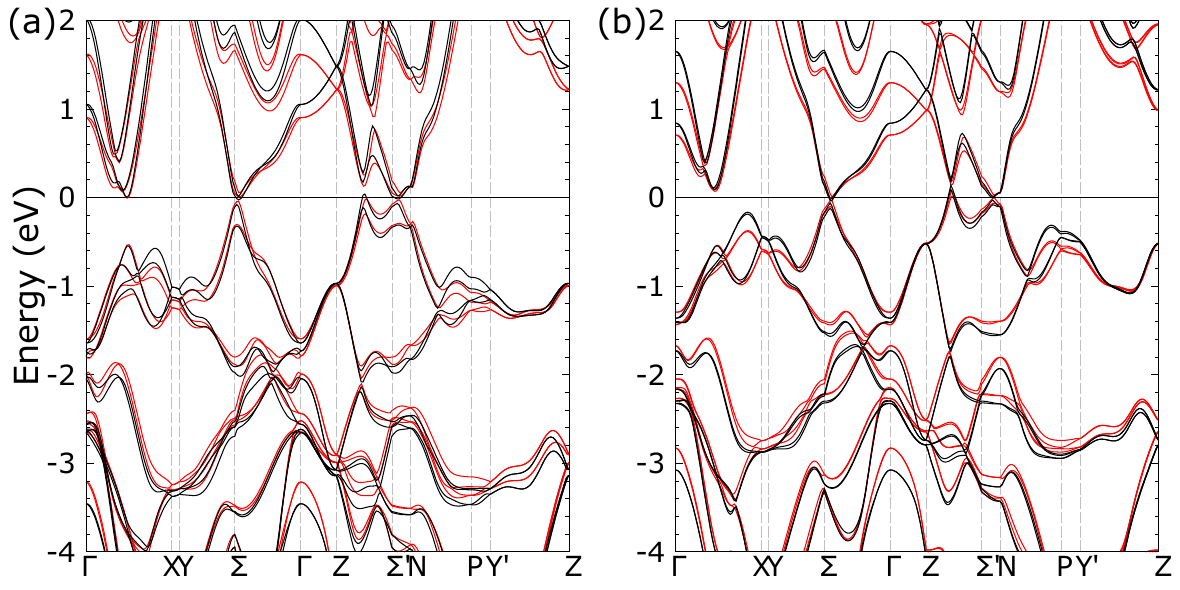}
    \caption{
      (Color online) Band structures of (a) TaX and (b) NbX (red line: X=As, black line: X=P) along high-symmetry lines in the bct BZ.
      }\label{SMfig:bands}
  \end{figure*}


  \begin{figure*}[h]
    \includegraphics[width=0.90\textwidth,height=0.95\textheight,keepaspectratio]{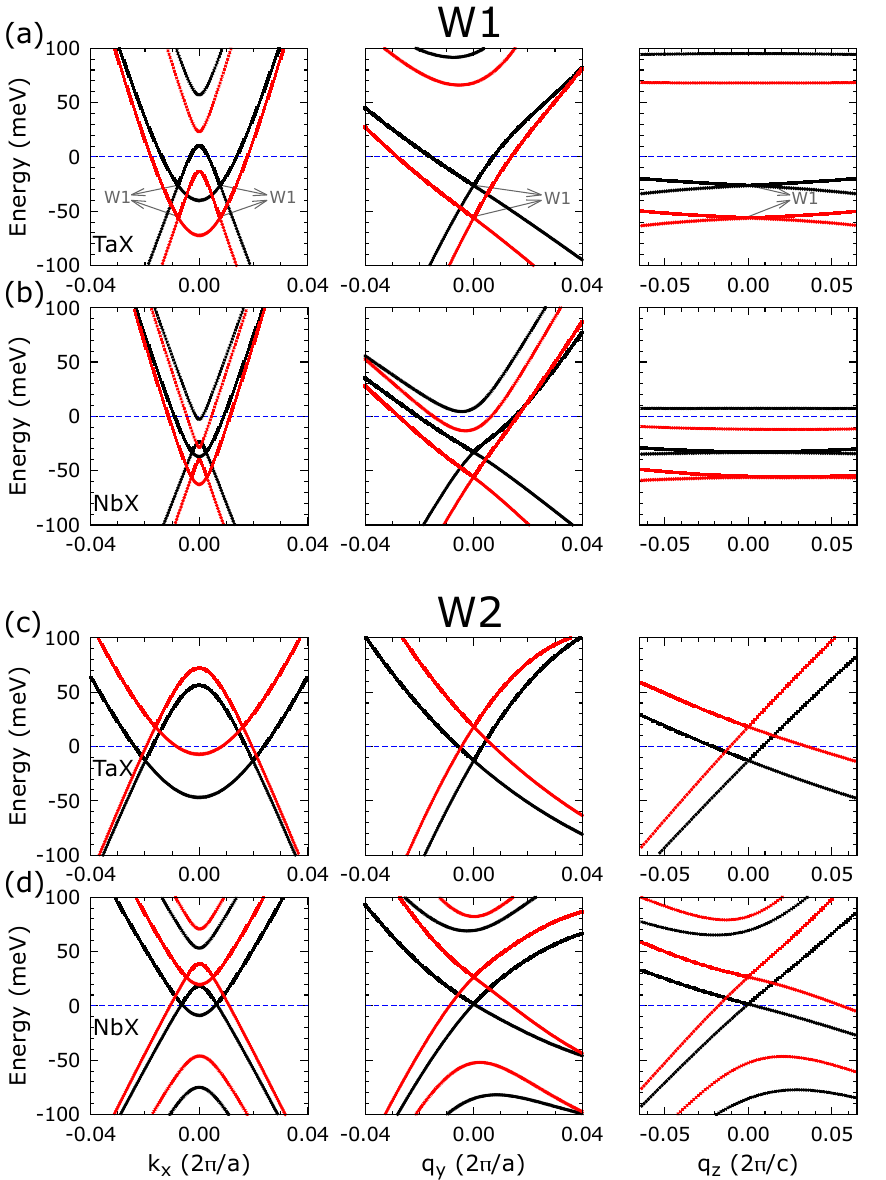}
    \caption{\label{SMfig:weyl}
      (Color online) Bands near the WPs at $\pm k_{Wx}$ with positive coordinates $k_{Wy}$ and $k_{Wz}$ in Table~\ref{SM:tab1} in a small energy interval around the Fermi level (blue dashed horizontal line) versus $k_x$ and wave vector deviations $q_y=k_y-k_{Wy}$ and $q_z=k_z-k_{Wz}$.
      The Weyl bands of TaX (a,c) and NbX (b,d) are displayed as black lines (X=As) and red lines (X=P), respectively.
      }
  \end{figure*}
  \begin{figure}[h]
    \includegraphics[width=0.5\textwidth,height=0.95\textheight,keepaspectratio]{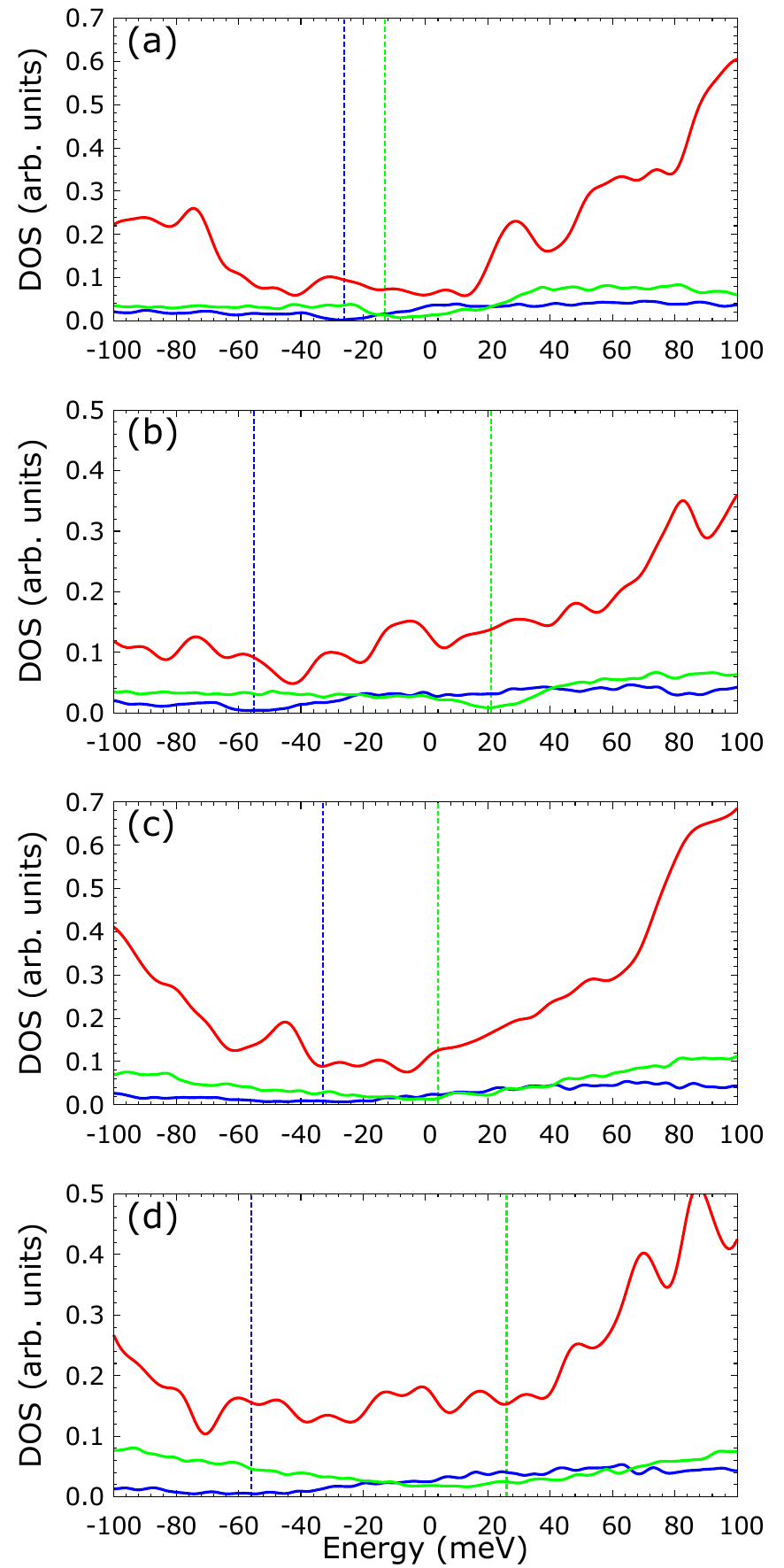}
    \caption{\label{SMfig:DOS}
      (Color online) Total density of states (red lines) of the four Weyl semimetals (a) TaAs, (b) TaP, (c) NbAs, and (d) NbP for small single-particle excitation energies including the trivial point contributions.
      Since the W1 and W2 Weyl nodes possess different absolute positions, the Fermi energy is used as energy zero.
      The positions of the Weyl nodes, where there are no states that contribute to the DOS, are given by $\varepsilon_{Fj}$ ($j=1,2$) and marked by blue (green) dashed lines for $j=1(2)$.
      The W1 (W2) contributions are displayed by blue (green) solid lines.
      }
  \end{figure}
  
   \begin{table}[h!]
    \caption{
      Weyl node positions in the Brillouin zone.
      }\label{SM:tab1}
    \begin{tabular}{ | c | r r r | r r r | }    \hline
      \multirow{2}{*}{}    & \multicolumn{3}{c|}{TaAs} & \multicolumn{3}{c|}{TaP} \\
                           & $k_x(2\pi/a)$ & $k_y(2\pi/a)$ & $k_z(2\pi/c)$ & $k_x(2\pi/a)$ & $k_y(2\pi/a)$ & $k_z(2\pi/c)$ \\ \hline
      \multirow{8}{*}{W1}  & 0.0078  & 0.5103  & 0.0000  & 0.0077  & 0.5174  & 0.0000   \\
                           & -0.0078 & 0.5103  & 0.0000  & -0.0077 & 0.5174  & 0.0000   \\
                           & -0.5103 & 0.0078  & 0.0000  & -0.5174 & 0.0077  & 0.0000   \\
                           & -0.5103 & -0.0078 & 0.0000  & -0.5174 & -0.0077 & 0.0000   \\
                           & 0.0078  & -0.5103 & 0.0000  & 0.0077  & -0.5174 & 0.0000   \\
                           & -0.0078 & -0.5103 & 0.0000  & -0.0077 & -0.5174 & 0.0000   \\
                           & 0.5103  & 0.0078  & 0.0000  & 0.5174  & 0.0077  & 0.0000   \\
                           & 0.5103  & -0.0078 & 0.0000  & 0.5174  & -0.0077 & 0.0000   \\ \hline
      \multirow{16}{*}{W2} & 0.0198  & 0.2818  & 0.5905  & 0.0161  & -0.2741 & 0.5885   \\
                           & -0.0198 & 0.2818  & 0.5905  & -0.0161 & -0.2741 & 0.5885   \\
                           & -0.2818 & 0.0198  & 0.5905  & 0.2741  & 0.0161  & 0.5885   \\
                           & -0.2818 & -0.0198 & 0.5905  & 0.2741  & -0.0161 & 0.5885   \\
                           & 0.0198  & -0.2818 & 0.5905  & 0.0161  & 0.2741  & 0.5885   \\
                           & -0.0198 & -0.2818 & 0.5905  & -0.0161 & 0.2741  & 0.5885   \\
                           & 0.2818  & 0.0198  & 0.5905  & -0.2741 & 0.0161  & 0.5885   \\
                           & 0.2818  & -0.0198 & 0.5905  & -0.2741 & -0.0161 & 0.5885   \\
                           & 0.0198  & 0.2818  & -0.5905 & 0.0161  & -0.2741 & -0.5885 \\
                           & -0.0198 & 0.2818  & -0.5905 & -0.0161 & -0.2741 & -0.5885 \\
                           & -0.2818 & 0.0198  & -0.5905 & 0.2741  & 0.0161  & -0.5885 \\
                           & -0.2818 & -0.0198 & -0.5905 & 0.2741  & -0.0161 & -0.5885 \\
                           & 0.0198  & -0.2818 & -0.5905 & 0.0161  & 0.2741  & -0.5885 \\
                           & -0.0198 & -0.2818 & -0.5905 & -0.0161 & 0.2741  & -0.5885 \\
                           & 0.2818  & 0.0198  & -0.5905 & -0.2741 & 0.0161  & -0.5885 \\
                           & 0.2818  & -0.0198 & -0.5905 & -0.2741 & -0.0161 & -0.5885  \\ \hline
      \multirow{2}{*}{}    & \multicolumn{3}{c|}{NbAs} & \multicolumn{3}{c|}{NbP} \\
                     & $k_x(2\pi/a)$ & $k_y(2\pi/a)$ & $k_z(2\pi/c)$ & $k_x(2\pi/a)$ & $k_y(2\pi/a)$ & $k_z(2\pi/c)$ \\ \hline
      \multirow{8}{*}{W1}  & 0.0026  & 0.4859  & 0.0000  & 0.0029  & 0.4921  & 0.0000 \\
                           & -0.0026 & 0.4859  & 0.0000  & -0.0029 & 0.4921  & 0.0000 \\
                           & -0.4859 & 0.0026  & 0.0000  & -0.4921 & 0.0029  & 0.0000 \\
                           & -0.4859 & -0.0026 & 0.0000  & -0.4921 & -0.0029 & 0.0000 \\
                           & 0.0026  & -0.4859 & 0.0000  & 0.0029  & -0.4921 & 0.0000 \\
                           & -0.0026 & -0.4859 & 0.0000  & -0.0029 & -0.4921 & 0.0000 \\
                           & 0.4859  & 0.0026  & 0.0000  & 0.4921  & 0.0029  & 0.0000 \\
                           & 0.4859  & -0.0026 & 0.0000  & 0.4921  & -0.0029 & 0.0000 \\ \hline
      \multirow{16}{*}{W2} & 0.0064  & 0.2790  & 0.5736  & 0.0047  & 0.2710  & 0.5770 \\
                           & -0.0064 & 0.2790  & 0.5736  & -0.0047 & 0.2710  & 0.5770 \\
                           & -0.2790 & 0.0064  & 0.5736  & -0.2710 & 0.0047  & 0.5770 \\
                           & -0.2790 & -0.0064 & 0.5736  & -0.2710 & -0.0047 & 0.5770 \\
                           & 0.0064  & -0.2790 & 0.5736  & 0.0047  & -0.2710 & 0.5770 \\
                           & -0.0064 & -0.2790 & 0.5736  & -0.0047 & -0.2710 & 0.5770 \\
                           & 0.2790  & 0.0064  & 0.5736  & 0.2710  & 0.0047  & 0.5770 \\
                           & 0.2790  & -0.0064 & 0.5736  & 0.2710  & -0.0047 & 0.5770 \\
                           & 0.0064  & 0.2790  & -0.5736 & 0.0047  & 0.2710  & -0.5770\\
                           & -0.0064 & 0.2790  & -0.5736 & -0.0047 & 0.2710  & -0.5770\\
                           & -0.2790 & 0.0064  & -0.5736 & -0.2710 & 0.0047  & -0.5770\\
                           & -0.2790 & -0.0064 & -0.5736 & -0.2710 & -0.0047 & -0.5770\\
                           & 0.0064  & -0.2790 & -0.5736 & 0.0047  & -0.2710 & -0.5770\\
                           & -0.0064 & -0.2790 & -0.5736 & -0.0047 & -0.2710 & -0.5770\\
                           & 0.2790  & 0.0064  & -0.5736 & 0.2710  & 0.0047  & -0.5770\\
                           & 0.2790  & -0.0064 & -0.5736 & 0.2710  & -0.0047 & -0.5770 \\ \hline
    \end{tabular}
  \end{table}
  \FloatBarrier
  
  \bibliography{bibliography}